\documentclass[12pt,preprint]{aastex}

\newcommand{\sgra}{Sgr~A$\textrm{*}$}

\renewcommand {\deg}   {\mbox{$^\circ$}}
\newcommand   {\arcm}  {\mbox{$^\prime$}}
\newcommand   {\arcs}  {\mbox{$^{\prime\prime}$}}

\newcommand   {\kms}   {\mbox{km\,s$^{-1}$}}
\renewcommand {\ga}    {\mbox{\rlap{\hbox{\lower5pt\hbox{$\sim$}}}\hbox{$>$}}}
\renewcommand {\la}    {\mbox{\rlap{\hbox{\lower5pt\hbox{$\sim$}}}\hbox{$<$}}}

\begin{document}
\pagenumbering{arabic} 
\def\kms {\hbox{km{\hskip0.1em}s$^{-1}$}} 
\voffset=-0.8in

\def\msol{\hbox{$\hbox{M}_\odot$}}
\def\lsol{\hbox{$\hbox{L}_\odot$}}
\def\kms{km s$^{-1}$}
\def\Blos{B$_{\rm los}$}

\def\psec           {$.\negthinspace^{s}$}
\def\pasec          {$.\negthinspace^{\prime\prime}$}
\def\pdeg           {$.\kern-.25em ^{^\circ}$}
\def\degree{\ifmmode{^\circ} \else{$^\circ$}\fi}
\def\ut #1 #2 { \, \textrm{#1}^{#2}} 
\def\u #1 { \, \textrm{#1}}          
\def\nH {n_\mathrm{H}}
\def\ddeg   {\hbox{$.\!\!^\circ$}}              
\def\deg    {$^{\circ}$}                        
\def\le     {$\leq$}                            
\def\sec    {$^{\rm s}$}                        
\def\msol   {\hbox{$M_\odot$}}                  
\def\i      {\hbox{\it I}}                      
\def\v      {\hbox{\it V}}                      
\def\dasec  {\hbox{$.\!\!^{\prime\prime}$}}     
\def\asec   {$^{\prime\prime}$}                 
\def\dasec  {\hbox{$.\!\!^{\prime\prime}$}}     
\def\dsec   {\hbox{$.\!\!^{\rm s}$}}            
\def\min    {$^{\rm m}$}                        
\def\hour   {$^{\rm h}$}                        
\def\amin   {$^{\prime}$}                       
\def\lsol{\, \hbox{$\hbox{L}_\odot$}}
\def\sec    {$^{\rm s}$}                        
\def\etal   {{\it et al.}}                     
\def\la{\lower.4ex\hbox{$\;\buildrel <\over{\scriptstyle\sim}\;$}}
\def\ga{\lower.4ex\hbox{$\;\buildrel >\over{\scriptstyle\sim}\;$}}
\def\refitem{\par\noindent\hangindent\parindent}
\oddsidemargin = 0pt \topmargin = 0pt \hoffset = 0mm \voffset = -17mm
\textwidth = 160mm  \textheight = 244mm
\parindent 0pt
\parskip 5pt

\shorttitle{Sgr A*}
\shortauthors{}

\title{Compact Radio Sources within 30\arcs\ of \sgra: Proper Motions,\\ 
Stellar Winds and the Accretion Rate onto \sgra}
\author{F. Yusef-Zadeh$^1$, H. Bushouse$^2$, R. Sch\"odel$^3$, M. Wardle$^4$, W. Cotton$^5$,\\ 
D. A. Roberts$^1$, F. Nogueras-Lara$^3$ \& E. Gallego-Cano$^3$
} 
\affil{$^1$Department of Physics and Astronomy and CIERA, Northwestern University, Evanston, IL 60208}
\affil{$^2$Space Telescope Science Institute, Baltimore, MD 21218}
\affil{$^3$Instituto de Astrofísica de Andalucía (CSIC), Glorieta de la Astronomia S/N, 18008 Granada}
\affil{$^4$Department of Physics and Astronomy, Macquarie University, Sydney NSW 2109, Australia}
\affil{$^5$National Radio Astronomy Observatory,  Charlottesville, VA 22903}

\begin{abstract} 
Recent broad-band 34 and 44 GHz radio continuum observations of the Galactic center have revealed 41 massive stars 
identified with near-IR counterparts, as well as 44 proplyd candidates within 30" of Sgr A*. Radio observations 
obtained in 2011 and 2014 have been used to derive proper motions of eight young stars near Sgr A*. The accuracy of 
proper motion estimates based on near-IR observations by Lu et al. and Paumard et al. have been investigated by using 
their proper motions to predict the 2014 epoch positions of near-IR stars and comparing the predicted positions with 
those of radio counterparts in the 2014 radio observations. Predicted positions from Lu et al. show an rms scatter of 
6 mas relative to the radio positions, while those from Paumard et al. show rms residuals of 20 mas, which is mainly 
due to uncertainties in the IR-based proper motions.
Under the assumption of homogeneous ionized winds, we 
also determine the mass-loss rates of 
11 radio stars, finding rates that are on average $\sim$2 times smaller than those 
determined  from model atmosphere calculations and near-IR data. 
Clumpiness of ionized winds would reduce 
the  mass loss rate of  WR and O stars 
by additional factors of 3  and 10, respectively. 
One important implication of this 
is a reduction in the expected mass accretion rate onto
\sgra\ from  stellar winds  by nearly an order of magnitude to a value of 
few$\times10^{-7}$ \msol\ yr$^{-1}$.   
Finally, we present the positions of 318 compact 34.5 GHz
radio sources within 30\arcs\ of \sgra. At least 45 of these have 
stellar counterparts in the near-IR $K_s$ (2.18 $\mu$m) and $L'$ (3.8$\mu$m) bands.  
The presence of a large number of compact radio sources suggests that 
high-frequency radio continuum observations of the Galactic center 
can not only characterize the properties of diffuse ionized gas, but also 
reveal the massive OB and WR stars, low and high mass young stellar objects 
and nonthermal sources, such as the magnetar 
SGR 1745-29, located within one parsec of \sgra. 
\end{abstract}

\section{Introduction}
The nuclear region of our Galaxy coincides with a stellar cluster consisting of a mixture 
of an evolved stellar population and a young stellar population centered on the 
4$\times10^6$ \msol\ black hole \sgra\ (Reid and Brunthaler 2004; Ghez \etal\ 2005; Gillessen \etal\ 2009). The 
young stellar cluster lies mainly within 1 to  10\arcs\ (0.04 to 0.4 pc) of the strong radio source 
\sgra, whereas the evolved cluster has a half-light radius of 1.8\arcm\ ($\sim$4.2 pc) centered on
\sgra\ (Sch\"odel \etal\ 2014; Fritz \etal\ 2014). The young population consists of about one 
hundred young massive OB and WR stars, which are contained within one 
or two disk-like distributions (Paumard \etal\ 2006; Lu \etal\ 2009).

Until recently, young stars in the two stellar disks could only be 
identified and studied  using adaptive optic observations  in the near-IR (NIR).
These observations, however, are not able to detect a steady component 
of the emission from the black hole. Thus the relative position of stars with respect to the 
stationary black hole can not be determined directly. 
Radio images, on the other hand, show 
the bright radio source \sgra\ and the ionized gas orbiting the black hole.  
The combination of radio and NIR observations  are therefore needed to make accurate astrometric measurements
of the various nuclear populations. 
The key  question that accurate astrometry can address 
is to make accurate  determinations of the positions, and hence 
reducing the errors and uncertainties in the positions of 
NIR-identified stars, such as S2, as they orbit \sgra.  
S2 has been observed over two decades and lies $\sim$0.2\arcs\ from \sgra\ with 
an orbital period of 15.8 years. 
A prograde periastron shift of 0.2$^\circ$ per one revolution around \sgra\ is estimated to yield a 
positional shift of 
$\sim0.8$ mas with respect to the previous orbit (Rubilar and Eckart 2001). Thus,
 the accurate position of S2 with respect to \sgra\ has important 
implications for probing the curvature of space-time in the context of general relativity. 

To compare radio and NIR images, the accurate position of the black hole with respect to stars is determined by 
aligning radio and NIR frames using SiO masers. These masers are associated with members of the evolved stellar 
cluster and are distributed more than 7\arcs\ from \sgra\ (Menten \etal\ 1997; Reid \etal\ 2003). We recently 
reported  detection of 41 radio sources (Yusef-Zadeh \etal\ 2014a).  
A comparison with NIR images of similar resolution 
taken at 
the same epoch identified all  radio sources having  L-band (3.8 $\mu$m)  counterparts (Yusef-Zadeh \etal\ 2014a).
Several young, massive stars of the central cluster within a few arcseconds of 
Sgr A*, such as IRS 16C, IRS 16NE, IRS 16SE2, IRS 16NW, 
IRS~16SW, AF, AFNW, IRS 34W and IRS 33E, are detected in the 44 GHz observations.
Most of these massive stars  show
P-Cygni profiles, which is strong evidence for stellar winds (Martins \etal\ 2007). 
We determined the mass-loss
rates of individual radio stars (Yusef-Zadeh \etal\ 2014a) 
assuming the standard model for a spherically-symmetric, homogeneous 
wind of fully
ionized gas (Panagia \& Felli 1975). 
The mass-loss rates obtained using the radio
and NIR techniques agree within a factor of
two. However, the mass-loss rate from IRS 33E  was estimated to  
have only one-third of that derived from model atmosphere calculations 
given by Martins \etal\ (2007).


Residual instrumental distortions, as well as a number of other uncertainties, in the mosaiced
results of NIR adaptive optics observations can be significant within 
10\arcs\ of \sgra\ (Fritz \etal\ 2010; Yelda \etal\ 2010). 
The detection of  young stars at radio wavelengths, however,  provides a
means of registering the Galactic center  radio and NIR frames to few milliarcsecond (mas) precision 
with respect to the position of Sgr A*.  
Radio counterparts of eight NIR stars within $\sim$10\arcs\ of \sgra\ have been identified.
The nearest  detected radio star, IRS 16NE, is only 1.1\arcs\ away from \sgra. 
Radio emission arises from the ionized winds close to 
the photospheres of massive stars, whereas SiO masers arise from  
inside the  dust formation zone of AGB stars. 
Localizing the radio emission from massive stars could therefore yield more accurate stellar positions than
the techniques that rely on SiO masers as a reference. 
Thus, accurate astrometry and proper motion of  sources near \sgra\ can 
be achieved by registering radio and NIR frames. 
Accurate estimates 
of the stellar velocity  near \sgra\ should have implications on whether any of the disk stars are unbound (see Reid 
\etal\  2007) or place a limit on the unseen mass of stars near \sgra.

Proper motions of massive  stars orbiting \sgra\ can also be detected in the radio.
Although relative proper motions 
can be accurately determined in NIR images, 
the first of such measurements  
assumed an  isotropic evolved stellar 
cluster, having zero mean motion with respect to \sgra, thus 
removing the average motion of the stars. 
At radio wavelengths, 
the determination of the proper motions of massive stars is similar to SiO maser-based measurements
and   can be determined without making any assumptions about 
the spatial distribution of the cluster.  More recently, Yelda et al. (2010) and 
Gillessen et al. (2009) 
have used the radio astrometric system from the masers and do not 
use the assumption of zero rotation mean motion. 
The advantage  of using young, massive   stars over SiO sources  is
that radio stars are  members of the young stellar cluster and lie within a few
arcseconds of \sgra, 
whereas SiO masers are more than  7\arcs\ away. 
Another advantage of measuring  proper motions in radio 
is that source confusion is far lower than in the NIR, reducing 
the  possibility of determining the positions of misidentified  stars.

Here we present the results of our recent sensitive measurements within the inner 30\arcs\ of \sgra.
First, in $\S3.1$, we  report the first proper motion measurements of  eight young stars 
determined using two epochs of radio observations separated by $\sim$31 months. 
These  eight  stars were detected in two epochs: 5 August 2011 
(2011.59) and 21 February 2014 (2014.14) 
Because there were no NIR observations at the same epochs as the radio observations,
we predicted the positions of NIR-identified stars
at the 2014.14 epoch and  compared  them to the radio positions. 
The predicted NIR positions were calculated using the proper motion 
measurements of Lu \etal\ (2009) and  Paumard \etal\ (2006).
Second, in $\S3.2$ we present updated mass-loss rates of massive stars from ionized stellar winds
at 34 and 44 GHz and compare them with the determination of  mass loss rates 
from NIR observations. 
Third,  in $\S3.3$ we give new and updated positions of radio stars near the Galactic center.
The list of 318 compact radio sources in Sgr A West 
includes the magnetar SGR 1745-29, detected at 34~GHz, 
and the identities of NIR  stellar counterparts to 45 of the sources.

\section{Observations and Data Reduction}
\subsection{Radio Data}

We obtained two sets of A-array observations with  the Karl G. Jansky Very Large Array       
(VLA). The first set of observations of the stellar cluster at the Galactic center were obtained on 
July 8--9, 2011 and August 31--September 1, 2011 at 44 GHz (program 11A-224).
The two pairs of A-array  observations  provided data for construction of  an image of the 
30\arcs\ region surrounding \sgra\, with a resolution of $\sim82\times42$ mas. 
Further details  of these first-epoch observations are given in Yusef-Zadeh \etal\  (2014a).
This yields a first epoch observation with an effective
date of 5 August 2011.  

We reobserved the region within 30\arcs\ of \sgra\ on 2014 February 21, again using the VLA in its A-configuration at 
44 GHz. The Q-band (7mm band) was used in the 3-bit system, which provided full polarization correlations in 4 
basebands, each 2 GHz wide, centered at 41.6, 43.6, 45.6 and 47.6 GHz, respectively.  Each baseband was composed of 16 
subbands, each 128 MHz wide.  Each subband was made up of 64 channels and channels were 2 MHz wide.
These two epochs of 44 GHz observations, with effective bandwidths of  
2$\times128$ MHz and 8 GHz, were used to measure proper motions. 

We 
also carried out A-array observations (program 14A-232) in the Ka (9mm) band on March 9,   2014 at 34.5 GHz.
We used  the 3-bit system, which provided full polarization in 4 basebands, each 2 GHz wide.
Each subband was made up of 64 channels and channels were 2 MHz wide.  We used the same 
calibration strategy as we did for the Q-band observations. 
In all our observations, we used 3C286 to 
calibrate the flux density scale and used 3C286 and J1733-1304 (aka NRAO530) to calibrate the bandpass.  
We used  J1744-3116 to calibrate the complex gains.  A Phase and amplitude self-calibration procedure was applied to 
all data using the bright radio source Sgr A*. 
We used OBIT (Cotton 2008)  to construct radio images. 
The positions of radio sources are  determined with respect to the absolute position of Sgr A*. 
Sgr A* provides a great
astrometric reference, especially at  high resolutions where 
there is very good contrast with diffuse  emission associated with orbiting ionized gas.  
The absolute position  and the proper motion of
Sgr A* are  known at a level of 10 mas (Yusef-Zadeh. Choate \& Cotton 1999; Reid \etal\, 2003) and 
$\sim$1 mas yr$^{-1}$  (Reid \etal\, 2003).  This
source is used as the astrometric reference as it cancels many  of the
systematic errors arising  from the use of  external calibrators.  


\subsection{Near-IR Data}

We searched for NIR counterparts to compact radio sources using high-angular resolution adaptive optics 
(AO) assisted imaging observations acquired with the VLT/NACO \footnote{Based on observations made with ESO Telescopes at 
the La Silla or Paranal Observatories under programs ID 089.B-0503}. 
A~$K_s-$band (central wavelength 
$2.18\,\mu$m) image was obtained in a rectangular dither pattern on 12 September 2012, using the S27 camera 
with a pixel scale of $\sim0.027"$\,pixel$^{-1}$ and locking 
the AO on the 
supergiant IRS\,7, located about 5.5\arcs\ north of \sgra. $L' $-band observations were obtained during various 
observing runs between June 2012 and September 2012. The $L'$ data were acquired in speckle mode, i.e.\ with the AO 
system switched off. Details of the observations are given in Table 1.  In $L' $ we observed five 
fields with different pointing and depths. Field\,1 was centered on \sgra, and Fields 2--5 were offset by 
approximately 20\arcs\ to the northeast, southeast, southwest, and northwest, respectively.


Standard NIR image data reduction was applied as a first step: flat-field correction, sky subtraction, and 
interpolation of bad pixels. Finally, the images of individual pointings were combined into large mosaics. In the case of 
the $L'$ image, the speckle holography technique, as described in Sch\"odel et al. (2013), was applied to the 
thousands of  speckle frames across the FOV to create final high-Strehl images for each pointing. 
The number of frames varies across FOV and is 
given by N $\times$ NDIT in Table 1: 
34000 in field 1 and  12750 in Field 4.  
The  Strehl value is estimated $\sim60$\%. 
The images of each 
pointing were then combined into a large mosaic. Pupil tracking was used during 
the observations to improve the subtraction of the high and rapidly varying background in $L'$. Therefore the 
field-of-view (FOV) changed continuously and the individual exposures had to be de-rotated. The changing FOV resulted 
in a slightly smaller effective area of the mosaic. 


We used the PSF fitting  package {\it StarFinder} (Diolaiti et al. 2000) to detect stars and measure their 
pixel positions and flux. Photometric calibration in $K_s$ was done with the star IRS\,16C (Rafelski et al. 2007, 
Sch\"odel 2010). From comparison with ESO zero points taken during routine quality control within a month of 
the observations we estimate a systematic uncertainty of our calibration of $1\,\sigma<0.05$\,mag. Photometric 
calibration in $L'$ was done with the stars IRS\,16C and IRS 16\,NW (Sch\"odel et al. 2010). Comparison with ESO zero 
points from June to September of the same year lead us to estimate a $1\,\sigma$ systematic uncertainty of about 
$0.1$\,mag. Because of the large field-of-view we extracted PSFs locally, for small overlapping fields, very similar 
to the procedure described in Sch\"odel (2010). From the difference of stellar fluxes measured with different PSFs in 
overlapping sub-fields we estimated the systematic uncertainty of the photometry due to limitations of our knowledge 
of the PSF. This uncertainty was estimated to be $0.05$\,mag for both $K_s$ and $L'$ and was added in quadrature to the 
formal uncertainties computed by {\it StarFinder}.
Finally, we calibrated the images astrometrically by using the positions and proper motions of SiO maser stars in the 
Galactic center (Reid \etal\ 2007). 
We used the positions and proper motions of the SiO masers IRS9, IRS7, IRS12N, IRS28, SiO-15, IRS10EE, IRS15NE, IRS17,
IRS19NW as published by (Reid et al. 2007) to calculate their position for the epoch of the images.  The masers' pixel
positions in the NIR images were measured via PSF fitting with StarFinder. Finally, the astrometry was solved with the
IDL solve-astro routine from ASTROLIB. NO distortion solution was fitted, just linear terms.
Note that we did not correct for any camera distortions. 
In order to check the astrometric accuracy we performed 100 runs of a Monte Carlo 
simulation. The maser positions were varied
randomly within their uncertainties (combined radio + NIR), using a normal distribution with a standard deviation
corresponding to the radio/NIR uncertainties. Each time the astrometry was solved and the NIR maser positions were
compared to their radio positions. From these simulations we obtained a standard deviation for the position of each
maser star. Additionally, we performed a jackknife test by using only 8 out of 9 maser stars. The standard deviations 
between the
(astrometric) NIR position and the prediction from the radio measurements were very similar to the Monte Carlo test.
From Monte Carlo 
simulations (varying the positions of the masers within their NIR and radio uncertainties),  we estimated that 
the astrometric precision  was better than one pixel (27 mas) across the region of the $K_s$ and $L'$ mosaics where the radio 
sources are located. 

\section{Results}
\subsection {Proper Motion Measurements}

We used two techniques to  compare radio and NIR proper motion measurements.
First, we measured the proper motions of the radio stars using data taken in 2011.59 (first epoch) and 2014.14 (second epoch). 
These estimates are made only from radio observations. 
We then compare these radio-derived proper motion values with those exclusively from NIR observations. 
A second technique 
involves the prediction of the NIR stellar source positions at the 2014.14 epoch and 
then comparing  those predicted positions with the 2014.14 radio positions.
The reason for doing this second type of comparison is because  we don't have
simultaneous radio and NIR observations. 


\subsubsection {Proper Motions of Radio Stars}

In our first and second  epoch radio observations on 2011.59 and 2014.14, the well-known young, 
massive stars in the central \sgra\ cluster (e.g., IRS 16C, IRS 16NE, 
IRS 16SE2, IRS 16NW, IRS 16SW, AF, AFNW, IRS 34W and IRS 33E),
as well as the  heavily extincted NIR source IRS 5, were detected 
with rms noise $\sigma \sim$61  (S/N ~ 4-20)
and $\sim22\, \mu$Jy (S/N ~ 4-31), respectively. 
These radio stars are isolated 
from the mini-spiral HII region and appear to be unresolved in two-dimensional Gaussian fits. 
Our analysis of the proper motion between August 2011 and February 2014 indicates
uncertainties that are dominated by the poor sensitivity of our 2011 data,
which suffered from relatively narrow bandwidths (2$\times$128 MHz). We measure 
proper motions of these eight stars with respect to \sgra. Columns 1 to  5  in  Table 2 
show the list of sources, the proper 
motion $\mu_x$ and $\mu_y$ in RA and   Dec, 
and the corresponding tangential velocity in RA and 
Dec, respectively.  A  proper motion of 1~mas yr$^{-1}$  corresponds to $\sim$40 \kms\ at the Galactic 
center distance of 8 kpc. 
As a reference, we also include in Table 2 the parameters of the fit 
to  \sgra\ (Reid and Brunthaler 2004).  
The assumed proper motion of Sgr A* is
6.379+/-0.024 mas yr$^{-1}$. This motion is due to 
the orbit  of the Sun around the center of the Galaxy (Reid and Brunthaler 2004).
The residual error left after removing  the solar motion is
$-0.4\pm0.9$ \kms\, running perpendicular to the plane of the Galaxy.
IRS 16C lies  
closest to \sgra\ in our sample and  gives  the highest proper motion values of
$-$11.82  and 9.8$\pm1.61$ mas yr$^{-1}$, in RA and Dec,  respectively.  We now compare these 
proper motion measurements with those determined from NIR data.


The orbits of young stars in the stellar disk are determined in the  NIR observations. This is generally 
done by measuring the proper motion of several hundred stars distributed around \sgra.  In this technique the zero 
point of the motion of the stellar cluster is determined by assuming isotropy. Unlike the NIR proper motion 
measurements, radio proper motion measurements (Reid \etal\,  2007), using either SiO masers or young radio stars, are 
carried out in the frame tied to \sgra, known to be stationary and requires no isotropy assumption. To compare 
our radio-derived proper motion values with those from NIR measurements, 
we list published measurements from Lu \etal\,  (2009) and Paumard
\etal\,  (2006) in Table 2.  The radio and NIR data generally agree within 1$\sigma$. The sources IRS 16C,  
AFNW,  and AF, however, show discrepancies between the two sets of measurements. IRS 16C, which is offset from \sgra\ by 
$\sim$1.17\arcs, is one of the brightest Galactic center massive stars in the central cluster, has a spectral type 
of Ofpe/WN9 (Martin \etal\,  2007), and significant tangential acceleration (Lu \etal\,  2009). 
The proper motion of IRS 16C 
deviates  by  $\sim3.3\sigma$  and $\sim1.2\sigma$ in RA and Dec when compared to NIR measurements 
given by Lu \etal\, (2009) and Paumard \etal\, (2006).  
The proper motions of IRS 16SE2, AFNW, and AF  differ by 
$\sim3.7\sigma$ in RA, $\sim2.3\sigma$ in RA, 
$\sim5.2\sigma$ in Dec  from those  of Paumard \etal\, (2006). 
The AFNW and AF stars (Paumard \etal\,  2006) are  isolated stars displaced from 
\sgra\ by $\sim$8.44\arcs\ and 9.47\arcs\ (sources 110 and 126 in Table 6), respectively.  
Although background subtraction is applied in our fitting 
procedures, these radio stars are  located in a region 
where there is very little nebular emission, so their flux and position measurements  
do not need background subtraction  and are  more accurate because 
they are not contaminated by confusing ionized features. 
The origin of these discrepancies is unknown, but it is possible that it 
is due to the curvature of the orbit that has not been accounted for in the IRS 16C proper motion 
measurements. Alternatively, it is possible that radio emission from  these stars 
arises not only from the ionized winds of mass-losing stars, but also 
from the bow shock due to the motion of the star with respect to the ISM near \sgra. 
The data presented here are  limited by the small sample of sources and high positional errors 
in our first epoch measurements. In spite of these limitations,
the radio and NIR  proper motion measurements  give consistent values for most 
of the stellar sources that we studied.  Future radio observations with similar sensitivity 
and resolution to those of our second  epoch observation should reduce the uncertainties in
radio-derived proper motions. 

 
\subsubsection{The Positions of Radio  and NIR--identified Stars}

Because we did not have  simultaneous observations of the stellar cluster at radio 
and NIR wavelengths, we were not able to   carry out direct
astrometric comparisons, as was done in 
Yusef-Zadeh \etal\,  (2014a). We therefore used a different approach for  the comparison of 
the positions of radio and NIR stars.

Using the catalogs of proper motion data given by  Paumard \etal\,  (2006) and Lu \etal\,  (2009)
we first computed the predicted X/Y offsets of NIR stars relative to \sgra, by evolving the orbits
of the stars to the 2014.14 epoch of radio observations. 
The 2014.14 positions were computed for a total of 31 and 90 stars from each of the above catalogs,
respectively.
These Right Ascension (RA) and Declination (Dec) offsets from \sgra\ were  then translated into 
absolute RA/Dec by assuming that the
RA/Dec (J2000) of \sgra\ is $17^h 45^m 40^s.0383,  -29^\circ 00'28''.069$.

The absolute coordinates of the orbit-evolved 
NIR sources 
were translated into the frame of our 2014.14 radio image and
sources common to both the NIR and radio catalogs were found using the IRAF task
``xyxymatch." We then 
determined a transformation between the radio and IR frames using the IRAF task ``geomap", which computes 
changes in 
scale factor and rotation between the coordinate systems of the two frames, in order to remove any overall 
differences between the radio and NIR frames 
due to instrumentation. This computation also gives us rms residual offsets 
between the two frames for the list of common sources.
 
A total of 6 radio stars in the 2014.14 radio image  were found to have NIR counterparts
in the catalog of objects from Lu \etal\, (2009).
The six identified  sources are IRS 16C, IRS 16NE, IRS 16NW, S3-5, IRS 16SW and IRS 33E.
Columns 1 to 7 of Table 3 lists the names of these six  stars, their predicted dX, dY offsets from 
\sgra\ in arcseconds and the corresponding errors X$_{err}$, Y$_{err}$,  and
the radio image pixel coordinates X(radio) and Y(radio), respectively. 
The predicted positions also include  the acceleration term given by 
Lu \etal\, (2009).  
\sgra\ is centered at pixel position 3456.0$\times$3456.0 in the
44 GHz radio image and    the pixel size is 8.68159 mas. 
The transformation of the two sets of coordinates between the radio and NIR images 
resulted  in a small overall shift of $\sim6$  and $\sim8$ mas in  the X and Y direction, respectively, 
which is at the level of 1$\sigma$ in the residuals of the overall transformation
There was no change in overall scale factor  and no indication of rotation to a level 
of $<0.35^\circ$. 
The rms residuals between the radio and NIR positions of the six sources are
on the order of 0.5--0.8 pixels in the X, Y directions,
which is equivalent to 5--7 mas in the radio image.

There were 21 NIR stellar sources from the catalog of Paumard \etal\, (2006) that had radio counterparts. 
We measured the corresponding pixel locations in the 44 GHz radio image 
for all of the NIR stars, which resulted in overall X and Y shifts between the two frames of 5        
and 1 mas, respectively, and again no indication of any significant change in either pixel scale or
rotation.
Columns 1 to 10 of Table 4 shows  the ID numbers and names of stars that Paumard et al. (2006) used, 
RA, Dec, the predicted offset positions from Sgr A*, the predicted NIR and radio 
 pixel positions in RA and Dec directions, respectively. 
The residuals in the positions of the 21 sources between the radio and NIR frames are
roughly 2.1 -- 2.6   pixels, 
equivalent to $\sim$18 to 22  mas in X and Y directions, respectively. 
The higher residuals in this set of measurements is mainly due to fairly large uncertainties in the proper motions 
that were used to extrapolate the positions of the NIR stars to the 2014.14 epoch of the radio data. Paumard \etal\, (2006)
do not quote uncertainties in the original positions of the NIR stars, so we assumed a typical uncertainty of 
$\sim$5 mas based on 
nature of their data. The average uncertainty in the proper motions leads to an additional scatter in the extrapolated 
2014 positions of $\sim$10 mas in each axis. When combined with the uncertainty of the original positions, this accounts 
for most of the $\sim$20 mas residuals in the positions of individual stars between the radio and NIR frames.


\subsection {Ionized Stellar Winds}

The radio stars that we detected at 44 GHz (Yusef-Zadeh \etal\,  2014a) 
coincide  with  luminous NIR sources having P-Cygni profiles, which is 
strong evidence for stellar winds  (Najarro \etal\, 1997; Martins \etal\, 2007).
The  radio  emission  most likely arises from  the  ionized winds 
of the hot stars (Panagia \& Felli 1975).
We previously determined the mass-loss rates of individual radio stars
from their 44 GHz emission (Yusef-Zadeh \etal\, 2014a) 
using  the standard model for a spherically-symmetric, homogeneous  wind of fully
ionized gas expanding with a constant terminal velocity (Panagia \& Felli 1975; 
Contreras \etal\,  1996). 
Given the improved signal-to-noise  ratio of the radio data that is presented here,  
we update the mass loss rates of  nine  well known stellar sources at  34.5  and 44.6 GHz. 
To estimate the mass loss rate, we use the expression 
by  Panagia and Felli (1975) given below, 

$$\dot{M} 
= 2.938\times10^{-6}\,\, 
 S_{\nu}^{0.75}\,\, \nu^{-0.6}\,\,  T_e^{-0.075}\, \mu\,\, v_{wind}\,\, d^{1.5}\, /Z^{0.5}  M_{\odot} yr^{-1},  $$

where  the flux density  S$_{\nu}$ is in  mJy, electron temperature 
$T_e$ in 10$^4$K, the mean molecular weight $\mu$ is  1.2, the
terminal wind velocity $v_{wind}$ in units of  10$^3$ \kms\,    the distance $d$ in  1 kpc and
the  solar metallicity Z=1
 For  S$_{\nu}=1$ mJy, $d$=8 kpc, and the above parameters, 
 stellar   mass loss rates are 
  $\dot{M} = 3.81\times10^{-5}$ and $3.39\times10^{-5}$  \msol\,  yr$^{-1}$  at 
 $\nu = 34.5$  and  44.6 GHz, respectively.

Table 5 compares  the mass-loss rates 
determined from the radio measurements at 34.5 GHz (8.695 mm) and 44.6 GHz (6.723 mm) with the 
NIR mass-loss rate estimates from model atmosphere calculations 
 (Martins \etal\, 2007). 
Columns 1 to 12 of Table 5 
show the NIR identified stellar sources, the terminal velocity, 
H to He ratio taken from Table 2 of Martins \etal\ (2007),  
the mean abundance,  the mass loss rate 
from NIR model atmosphere calculations, 
the peak flux density at 34.5 GHz, the mass-loss rate derived from 34.5 GHz measurements, 
the peak flux density at 44.6 GHz and the mass-loss rate derived from 44.6  GHz measurements, 
respectively. 
The mean metallicity was estimated for each source using H ans He abundances given 
by Martins \etal\ (2007). 
We include the mass-loss rate of four additional sources only at 34.5 GHz as well as 
IRS 16NE and  IRS 16SW using radio data, even though 
they are binaries (Pfuhl \etal\, 2014). The  model atmosphere calculations for  these two binaries 
are not available.  
Table 5 adopts  the Galactic center distance 7.62 kpc, as was used by Martins \etal\ (2007).  
A comparison between radio and NIR-determined
stellar  mass loss rates is shown in Figure 1. 
The filled red points  show stellar mass loss rates 
derived in the NIR and those derived from 34.5 GHz fluxes.   Similarly, 
the unfilled blue points  show the mass loss determined at 34.5 and 44.6 GHz. 
We note that radio-determined  mass-loss rates are systematically lower than
those of NIR-determined mass-loss rates.  
The  red dashed line shows 
the best-fitting line assuming that the NIR-derived mass-loss rate is 
directly proportional to the 
radio-derived rate whereas  the  blue points 
show mass loss rates derived from radio flux at 44.6  GHz vs that derived at 
34.5 GHz.  
The black dashed line indicates where the mass loss rates derived at different frequencies are equal.
The offset between the red and black dashed lines shows  
that radio-determined mass-loss rates are on average  lower 
than NIR-identified mass-loss rates by a factor of 2.

Radio observations provide the gross properties of the wind, whereas model atmosphere calculations require an estimate 
of the extinction, the detailed ionization balance of trace species, as well as the shape of the velocity profile. 
Thus, the mass-loss rate under the assumption of homogeneity and spherical symmetry using radio continuum data is 
generally simple and straight-forward to derive and perhaps more accurate (e.g., Blomme 2011). A number of recent 
studies indicate that there is overwhelming evidence for clumpy ionized winds from massive stars (e.g., review by 
Crowther 2007) and that the clumpiness factor reduces the mass-loss rate.
The  clumping factor $f_{cl}= < \rho^2 > /< \rho 
>^2$ where the brackets indicate an average density 
over the volume in which radio continuum at all wavelengths is formed 
(Blomme 2011). 
For example, clumpiness reduces the global mass-loss 
rates of WR stars by roughly a factor of 3 (Crowther 2007) 
whereas the winds from O stars, which are highly clumpy, reduces the 
mass-loss rate estimate by an order of magnitude or even more  (Crowther 2007; Sundqvist et al. 2011).


The mass-loss rates determined from the radio emission of Galactic center O and WR stars are systematically lower than 
that from NIR estimates by an average value of 2 for a total of 11 sources, as shown in Figure 1. The 
discrepancy in mass-loss rate estimates 
become more significant if the ionized winds are clumpy. We assume that the 11 sources in Table 5 represent a sample 
of mass-losing luminous stars in the Galactic center. Assuming WR stars only, the mass-loss rates from all stars are 
reduced roughly by a factor of 6. If we assume all stars are O-type, the mass-loss rate is reduced by a factor of 20. 
Given a mixture of WR and O stars distributed within 10$''$ of Sgr A*, the mass-loss rate estimates from OB and WR 
stars are reduced by roughly an order of magnitude. Numerical simulations of wind accretion onto Sgr A* show that the 
accretion rate is $\sim3\times10^{-6}$ \msol\ yr$^{-1}$ after accounting for the motion of mass-losing OB and WR stars 
in a disk geometry orbiting Sgr A* (Cuadra \etal\ 2006). The reduced estimate of mass-loss rates from OB and WR stars 
from our radio observations implies that the accretion rate onto Sgr A* should be roughly 3$\times10^{-7}$ \msol\ 
yr$^{-1}$. We note that our  reduced  estimate of the wind accretion rate is consistent with the mass accretion 
estimate made from polarization measurements (Bower et al. 2003).



One of the  implications of the  reduced mass-loss rate of luminous young stars 
is the reduction in the  Bondi-Hoyle  accretion rate onto \sgra. 
This is because
the current paradigm assumes  that \sgra\ is fed by the partial accretion of merged stellar winds 
from the OB and WR stars in the inner parsec of the Galaxy (Coker \&  Melia  1997;
Quataert 2004; Cuadra \etal\,  2006, 2008).  
The Bondi accretion rate onto \sgra\ will
be significantly reduced if the ionized winds are highly clumpy, thus
reducing the  expected luminosity of \sgra\  by more than an order of magnitude.




\subsection{Compact Radio Sources  at  34 GHz}

One of the most interesting aspects of the new broad-band radio continuum images of the Galactic center is the 
detection of numerous compact radio sources, many of which appear to coincide with massive stars, photoevaporative 
proplyd-like objects, or evaporative gaseous globule candidates (Yusef-Zadeh \etal\, 2014a, 2015). 
The improved sensitivity of these new images allows for the detection of isolated weakly emitting radio sources,
as well as resolving radio stars embedded within the diffuse ionized gas. 
Table 6 gives a list of 318 compact sources at 34.5 GHz distributed within
30\arcs\ of \sgra. Entries in columns 1 to 11 in Table 6 give the source ID, NIR identified stellar source, RA 
and Dec, the angular distance from \sgra, the positional precision, the deconvolved size, peak radio intensity, 
integrated intensity, and the NIR flux density at $K_s$ and $L'$ bands. 
Column 12 provides the references and comments on individual sources.

Identified NIR counterparts to the compact radio sources are listed column 2 of Table 6. Note that some of 
these stars may be variable, but we did not investigate this here. Sources detected in both $K_s$ and $L'$ can be 
considered high confidence detections. Also, the sources detected only at $K_s$ appeared to be clearly point-like, 
although surprisingly faint. 
It is not clear why  they appear so faint, perhaps  because of differential extinction
across the Galactic center, thus reducing the flux. Alternatively, they could be 
pre-main sequence stars photoionized externally by the strong radiation field (Yusef-Zadeh \etal\, 2015).
We identified NIR counterparts to the radio sources by simple positional matching.
Sources were accepted as identical if they coincided within two NACO pixels or about 54\,mas, which corresponds to the 
angular resolution of NACO in the $K_s-$band. The resolution of the radio image is $\sim47\times88$ mas. 
Because the radio and NIR data are not from the same epoch, the 
positions of the stars will not match  precisely. This is why we used a relatively large tolerance for matching 
the positions. Nevertheless, the tolerance is sufficiently small to exclude confusion with other stars.
Table 6 also shows that  at least 45 of these compact radio sources have 
stellar counterparts in the near-IR $K_s$ (2.18 $\mu$m) and $L'$ (3.8$\mu$m) bands.  
The number of stellar sources with radio counterparts is limited because   
the  spatial coverage of  the radio images is  $\sim1'\times1'$,  whereas the region  covered by the five $L'$ fields  
is $\sim50''\times50''$  and  there are gaps between the fields. 

The positions and the sizes  of radio sources are determined from background-subtracted Gaussian 
fits to the individual radio sources.  
Radio sources that are unresolved are shown by dashed lines.  Note the inclusion of the magnetar 
SGR~J1745-29, ID 13 in Table 6 (Kennea et al.  2013; Shannon and Johnston 2013).  
This  source was in its  quiescent phase 
 before it was identified as an X-ray outburst (Kennea et al. 2013). 
SGR J1745-29  is the closest known pulsar 
located 2.4\arcs\ from \sgra. The detection of a compact radio source located at $\alpha\, \delta\, (J2000) = 17^h 45^m 
40^s.16795\pm0.00002\, -29^{\circ} 00' 29''.74908\pm0.00064$ was reported at 44.6 GHz on 2014 February 21 
(Yusef-Zadeh \etal\, 2014b).  
Another notable collection of radio sources coincides with IRS 21, which  is comprised of 
six  radio components,  suggesting that IRS 21 is a stellar cluster similar to IRS 13N and IRS 13E. 


There are several radio sources that have NIR counterparts  only at $L'$ (Table 6). 
Several of these sources can be found within 0.1\,pc of \sgra. These sources
 may be similar in nature 
to the Dusty S-cluster Object DSO/G2 that is currently the subject of intense observational campaigns as it passes 
extremely close to the central black hole (Gillessen et al. 2012; Witzel et al. 2014; Pfuhl et al. 2015; 
Valencia-Schneider et al. 2015). Alternatively, 
recent radio study of $L'$ sources with radio counterparts argues  that  they are 
massive young stellar objects (YSOs). In this  interpretation, 
thermal radio emission  from the ionized gas  of the $L'$ sources is being photo-evaporated from  the disks
of YSOs by the UV radiation from hot stellar sources in the Galactic center (Yusef-Zadeh et al. 2014a).

Figure 2 shows a 34.5 GHz image of Sgr A West. A close-up view of this image  includes
circles and ID numbers, from column 1 of Table 6, identifying compact radio sources. 
Figure 3 shows a 3.8$\mu$m image of the region within $\sim20$\arcs\ of \sgra.
Labels show  the  NIR-identified stellar sources that have radio counterparts, 
as listed in column 2 of Table 6. The insets 
show close-up views of the crowded regions of Sgr A West  where 
the stellar clusters  IRS 16 and IRS 13 lie. 
Figures 4a,b show large and scaled-up views of the central 
region of Sgr A West at 34.5 GHz, respectively. Labels indicate radio stars with NIR stellar counterparts, 
as shown in 
Figure 3.


\section{Conclusions}

With the improved sensitivity of recent VLA radio observations of the Galactic center, we provided three aspects of 
the radio properties of objects within 30\arcs\ of \sgra\, as discussed below. 

First, we calculated  the proper motion of radio 
stars orbiting \sgra.  Our proper motion measurements illustrate that future high resolution radio observations at 
multiple epochs can compete with other methods of determining the orbit of stars, as done at NIR wavelengths. One main 
advantage that radio studies provide is that the positions of stars are measured directly and accurately with respect 
to \sgra, a feat that is not possible with NIR imaging.  This is because of the lack of a steady component of NIR 
emission from \sgra\ which anchors the radio frame. 
Another advantage is that source confusion is far lower  in the radio  than in the NIR, thus, one 
can avoid biased positions of the young stars at NIR. Lastly, the relative motion between the 
cluster and \sgra\ which has already been tested for SiO masers (Reid et al. 2007), but the greater number of radio 
stars may provide  useful constraints.

Second, we estimated the mass-loss rates of massive stars near \sgra\ using radio data and compared these values with 
those estimated from model atmosphere calculations. 
Assuming that the  winds  are clumpy,
the  discrepancy between radio and NIR mass-loss estimates 
would reduce the near-IR-determined mass loss rate of  WR and O stars
by factors ranging between  6  and 20, respectively.    
This implies a downward revision of the  mass accretion rate 
by an order of magnitude  to $\sim3\times10^{-7}$ \msol\ yr$^{-1}$ 
if the merging of ionized winds from massive stars is responsible for the accretion onto \sgra\ (Cuadra \etal\ 2006).

Third, we reported the detection of 318 radio sources within the inner 
30\arcs\ of \sgra. The comparison of radio and NIR data  indicated  that at least 45 of 
the radio sources are stars and have counterparts at both $K_s$ and $L'$ bands. 
The radio sources include compact HII regions, proplyd-like objects, externally heated massive YSOs, 
ionized winds from massive stars as well as
nonthermal sources. Future studies  of the spectra of these sources will be useful in 
identifying the true nature of radio sources within 30\arcs\ of Sgr A*. 
In addition,  the future VLTI instrument GRAVITY can provide astrometry with a precision up to 
10 microarcseconds, but only relative to a reference star within 2\arcs\ of the target. 
To  measure orbits of stars around Sgr A*, it is 
still necessary to tie them into the large astrometric reference frame. The detection of radio stars opens a new window 
for astrometric calibration. There are several radio stars within 2\arcs\ of  Sgr A* so that can be 
 used
 for calibration and for testing the accuracy of the calibration. Even if only a single reference star is used for GRAVITY 
(IRS 16C or IRS 16NW), then one can still compare the two calibration methods: (a) the traditional NIR-maser proper motion 
calibration and (b) a more direct calibration via radio detection and proper motions.





Acknowledgments:
This work is partially supported by the grant AST-0807400 from the NSF
and the European Research Council
under the European Union's Seventh Framework Program (FP/2007-2013). 
The National Radio Astronomy Observatory is a facility of the National Science Foundation, operated under a 
cooperative agreement by Associated Universities, Inc.

\begin{figure}
\center
\includegraphics[scale=0.8,angle=0]{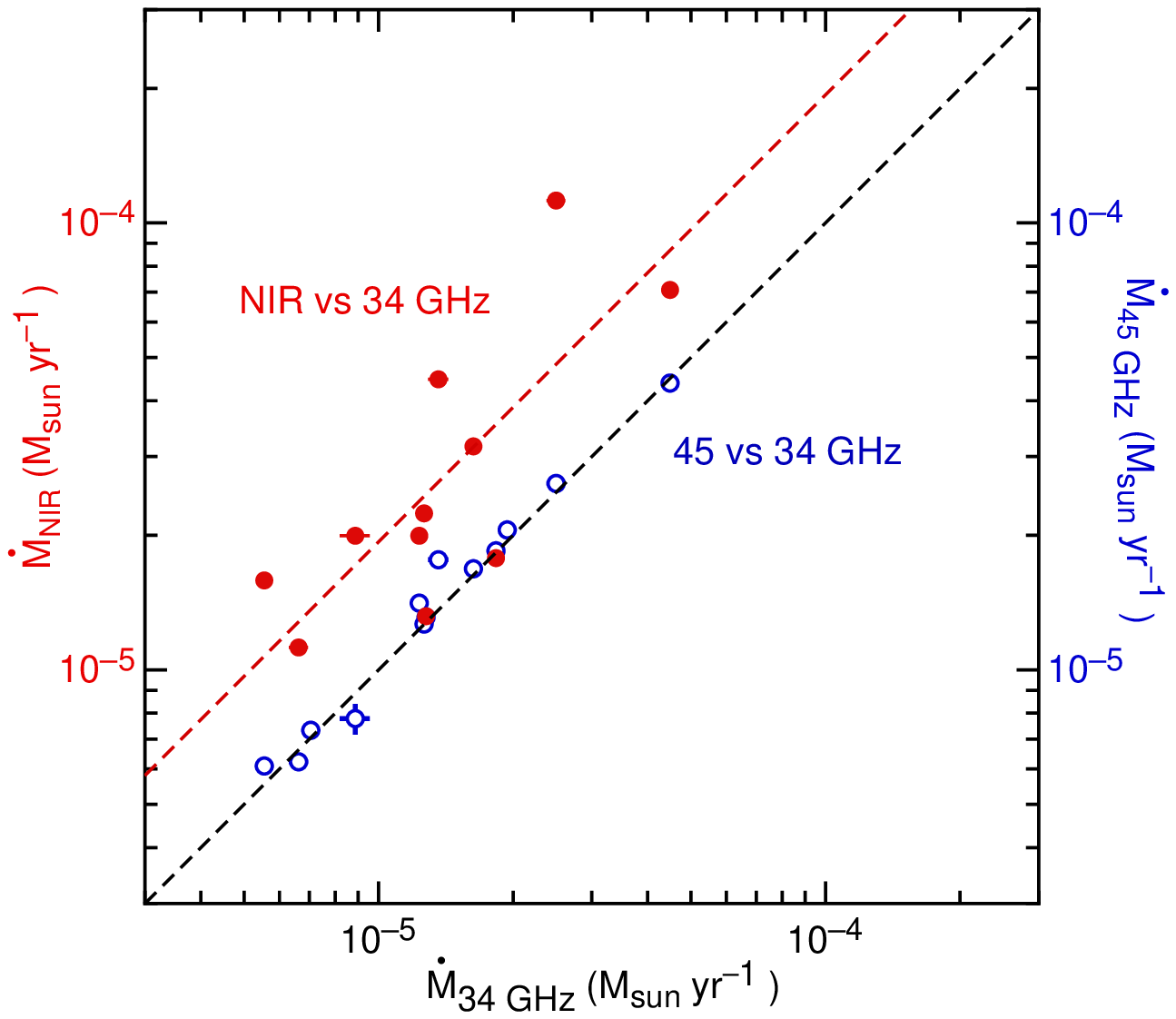}
\caption{ 
Comparison of stellar mass loss rates derived in the NIR with those derived from 34.5 GHz fluxes shown in red.  Red 
dashed line shows the best-fitting line to the red points (see text).  
Blue points show mass loss rate derived from radio flux at 44.6
GHz flux vs that derived at 34.5 GHz.  The black dashed line indicates
where the  mass loss rates derived at different frequencies are equal.
}
\end{figure}

\begin{figure}
\center
\includegraphics[scale=0.8,angle=0]{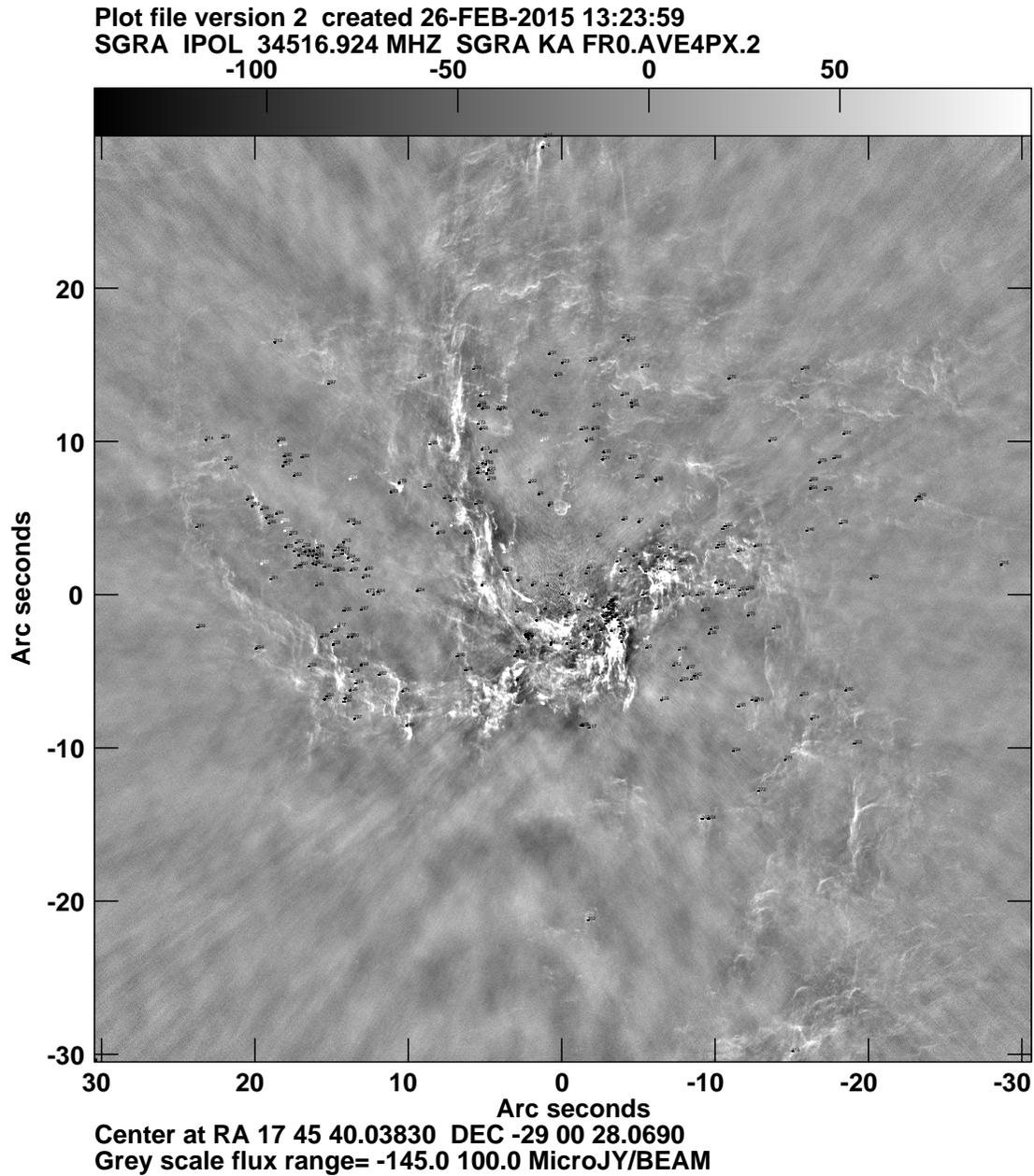}
\caption{
A 34.5 GHz image of Sgr A West 
with the  spatial resolution of  886$\times465$ mas with PA=$-1.6^\circ$. 
Every 4 pixels is displayed because of the large size of the image.  
Filled circles coincide with  the positions of 318 sources given in Table 6.
The labels  can be viewed when the image is 
enlarged.  (High resolution images in this paper are available 
by request). 
}
\end{figure}

\begin{figure}
\center
\includegraphics[scale=0.6,angle=0]{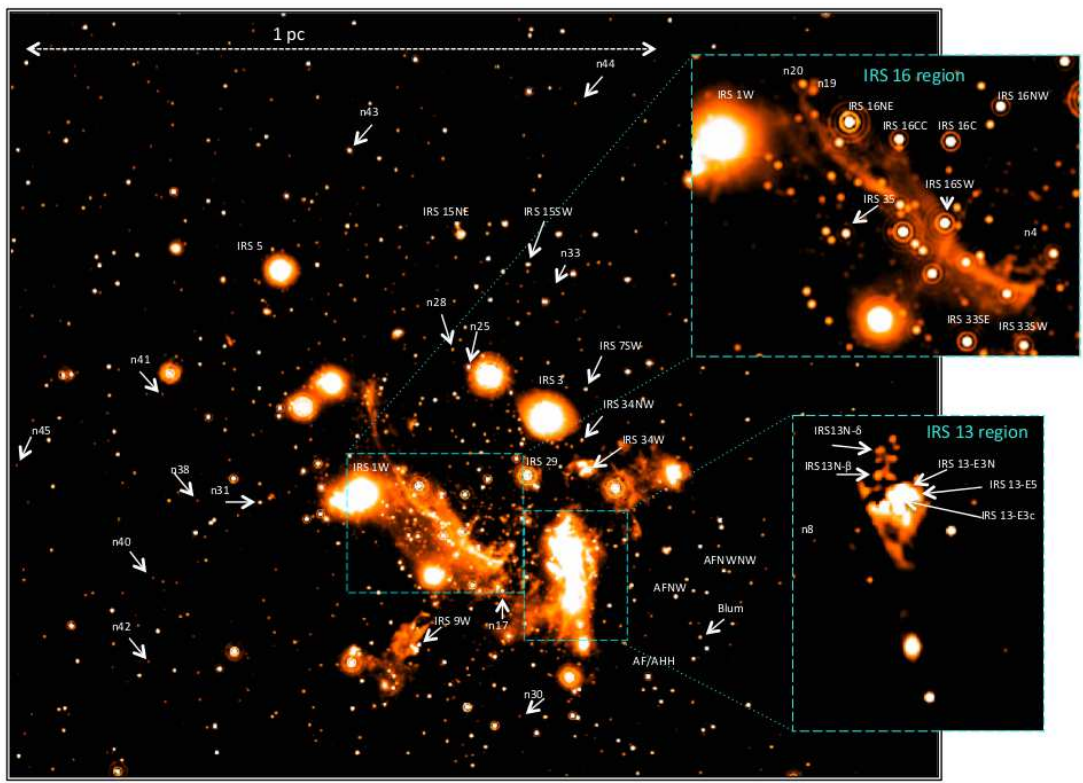}
\caption{
A 3.8$\mu$m image within  20$''\times20''$  of Sgr A*. 
Labels  show NIR sources with radio counterparts with positions given in Table 6. Insets show 
 close-up views of two stellar clusters IRS 16 and IRS 13.  
}
\end{figure}

\begin{figure}
\center
\includegraphics[scale=0.8,angle=0]{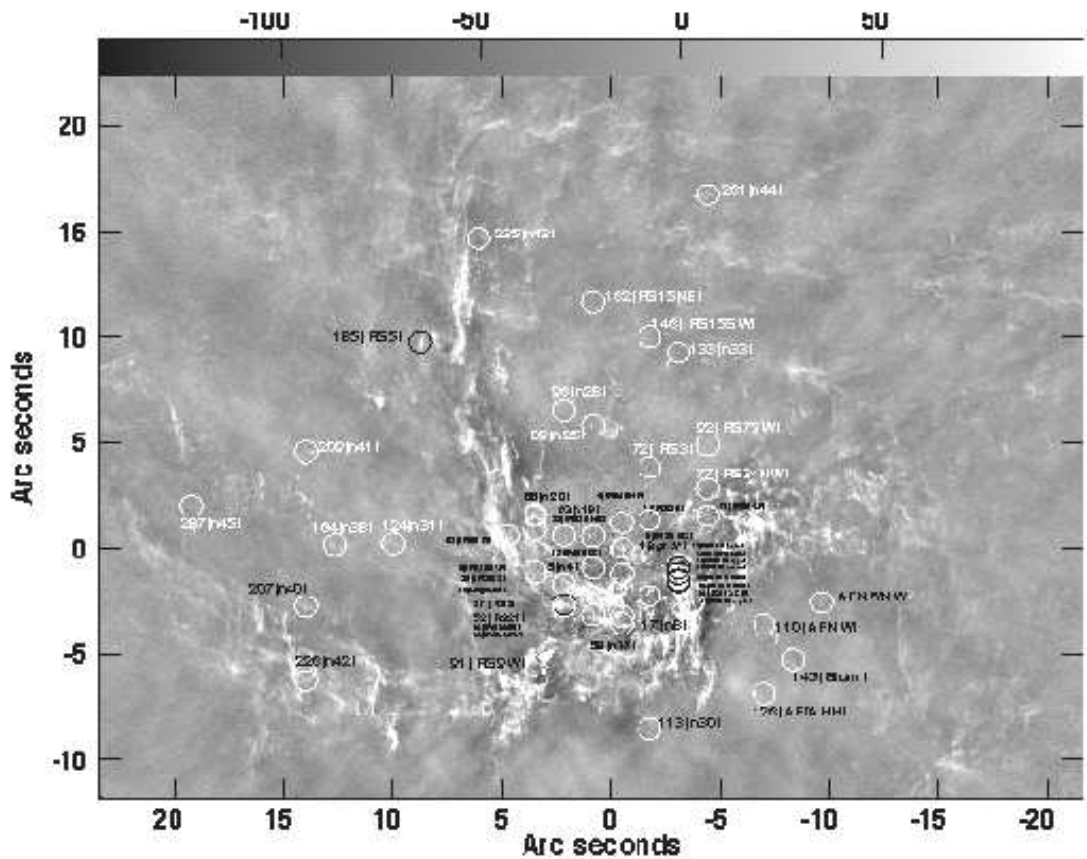}
\caption{{\it (a)}
Similar to Figure 1 except that the central 
$\sim40''\times30''$ of Sgr A West is shown. 
Circles  and labels point to the positions of radio sources with NIR counterparts. 
{\it (b)}
Similar to (a)  except that the central $\sim5''\times7''$
is displayed. Small labels are visible if zoomed in (as in Fig. 2). 
}
\end{figure}

\vfill\eject

\begin{figure}
\center
\includegraphics[scale=0.8,angle=0]{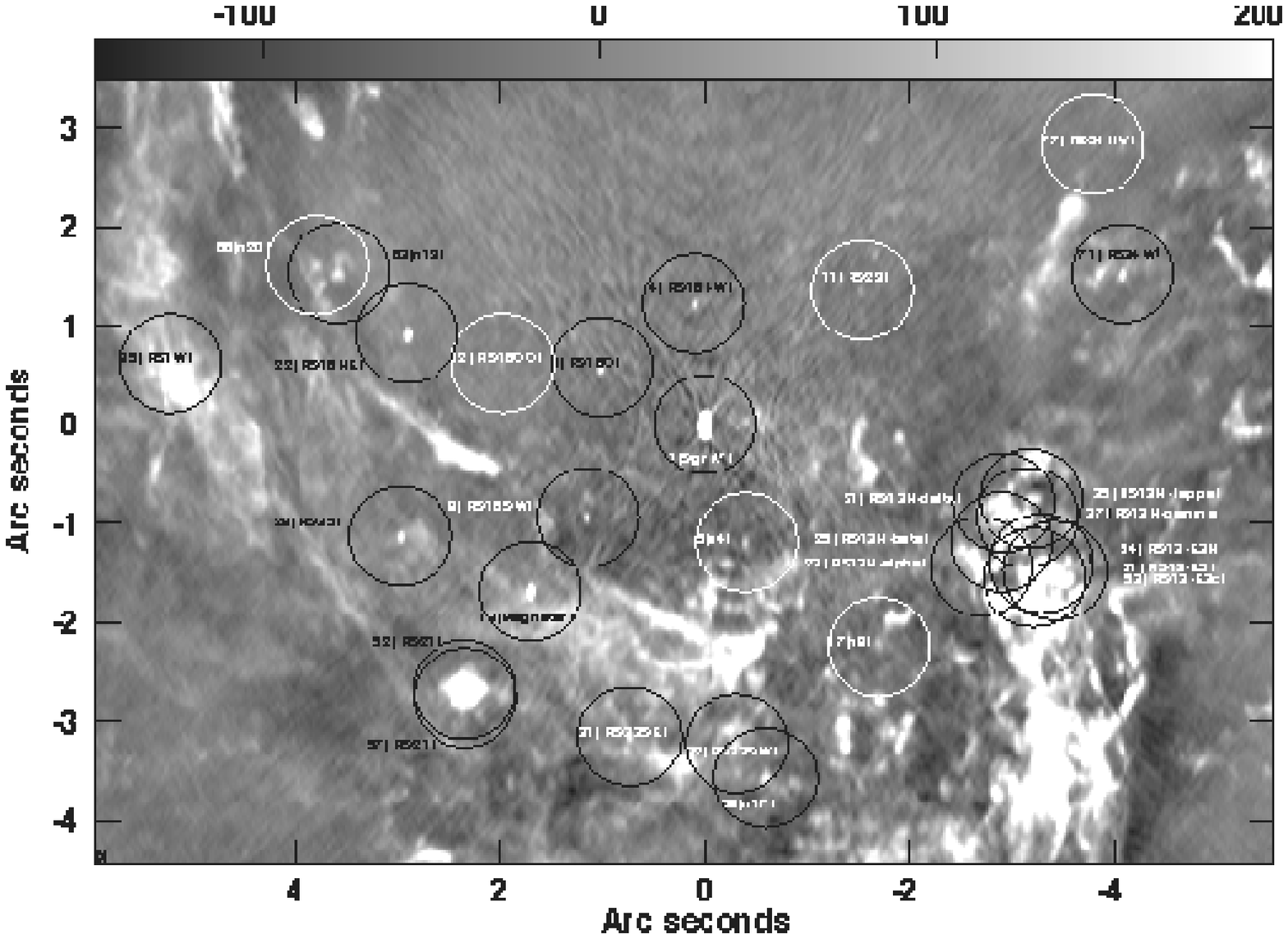}
\end{figure}

\begin{table}
\centering
\caption{Details of IR Observations}
\label{Tab:Obs}
\begin{tabular}{llllll}
\hline
\hline
Date & $\lambda_{\rm central}$ & $\Delta\lambda$ & N$^{\mathrm{a}}$ & NDIT$^{\mathrm{b}}$ & DIT$^{\mathrm{c}}$\\
 &  [$\mu$m]  &   [$\mu$m] &  & & [s] \\
\hline
12 Sept 2012 & 2.18 & 0.35 & 32 & 60 & 1 \\
\hline
$L'$ Field 1 & & & & & \\
27 June  2012 & 3.80 & 0.62 & 102 & 125 & 0.25 \\
14 July  2012 & 3.80 & 0.62 & 68 & 125 & 0.25 \\
20 July  2012 & 3.80 & 0.62 & 34 & 125 & 0.25 \\
12 Aug  2012 & 3.80 & 0.62 & 34 & 125 & 0.25 \\
14 Aug  2012 & 3.80 & 0.62 & 34 & 125 & 0.25 \\
\hline
$L'$ Field 2 & & & & & \\
17 July  2012 & 3.80 & 0.62 & 34 & 125 & 0.25 \\
20 July  2012 & 3.80 & 0.62 & 34 & 125 & 0.25 \\
12 Aug  2012 & 3.80 & 0.62 & 34 & 125 & 0.25 \\
14 Aug  2012 & 3.80 & 0.62 & 34 & 125 & 0.25 \\
\hline
$L'$ Field 3 & & & & & \\
17 July  2012 & 3.80 & 0.62 & 34 & 125 & 0.25 \\
20 July  2012 & 3.80 & 0.62 & 34 & 125 & 0.25 \\
12 Aug  2012 & 3.80 & 0.62 & 34 & 125 & 0.25 \\
26 Aug  2012 & 3.80 & 0.62 & 34 & 125 & 0.25 \\
6 Sep  2012 & 3.80 & 0.62 & 34 & 125 & 0.25 \\
\hline
$L'$ Field 4 & & & & & \\
17 July  2012 & 3.80 & 0.62 & 34 & 125 & 0.25 \\
20 July  2012 & 3.80 & 0.62 & 34 & 125 & 0.25 \\
13 Aug  2012 & 3.80 & 0.62 & 34 & 125 & 0.25 \\
\hline
$L'$ Field 5 & & & & & \\
17 July  2012 & 3.80 & 0.62 & 34 & 125 & 0.25 \\
2 Aug  2012 & 3.80 & 0.62 & 34 & 125 & 0.25 \\
13 Aug  2012 & 3.80 & 0.62 & 68 & 125 & 0.25 \\
21 Aug  2012 & 3.80 & 0.62 & 34 & 125 & 0.25 \\
26 Aug  2012 & 3.80 & 0.62 & 34 & 125 & 0.25 \\
\hline
\end{tabular}
\begin{list}{}{}
\item[$^{\mathrm{a}}$] Number of (dithered) exposures
\item[$^{\mathrm{b}}$] Number of integrations that were averaged on-line by the read-out
 electronics
\item[$^{\mathrm{c}}$] Detector integration time. The total integration time of each observation amounts
to N$\times$NDIT$\times$DIT
\end{list}
\end{table}

\begin{deluxetable}{lcccc}
\tablecaption{Proper motion of radio stars: Epoch1=2011.59, Epoch2=2014.14}
\tabletypesize{\scriptsize}
\tablecolumns{5}
\tablewidth{0pt}
\setlength{\tabcolsep}{0.04in}
\tablehead{
\colhead{Source} 
& \colhead{Proper Motion RA} & \colhead{Proper Motion Dec} & \colhead{Proper Motion RA} & 
\colhead{Proper Motion Dec}\\
& (\sl{mas yr$^{-1}$}) & (\sl{mas yr$^{-1}$}) & (\sl{km s$^{-1}$}) & (\sl{km s$^{-1}$}) 
}
\startdata
IRS 16NE (7mm) & $ 2.78 \pm 0.47 $ & $ -11.18 \pm 0.82 $ & $ 111.2 \pm 18.8 $ & $ -447.4 \pm 32.9 $\\
IRS 16NE (IR-Lu) & $ 3.11 \pm 0.06 $ & $ -10.94 \pm 0.06 $ & ----- & ----- \\
IRS 16NE (IR-Paumard) & -----  & ----- & $ 104 \pm 49 $ & $ -379 \pm 47 $\\
\\
IRS 16C (7mm) & $ -12.03 \pm 0.99 $ & $ 9.64 \pm 1.65 $ & $ -481.2 \pm 39.5 $ & $ 385.7 \pm 65.9 $\\
IRS 16C (IR-Lu) & $ -8.74 \pm 0.05 $ & $ 7.42 \pm 0.05 $    & -----   & -----  \\
IRS 16C (IR-Paumard) & -----  & ----- & $ -342 \pm 50 $ & $ 302 \pm 44 $\\
\\
IRS 16NW (7mm)  & $ 7.13 \pm 1.74 $ & $ -3.46 \pm 3.12 $ & $ 285.4 \pm 69.8 $ & $ -136.6 \pm 124.9 $\\
IRS 16NW (IR-Lu)  & $ 6.30 \pm 0.06 $ & $ 0.87 \pm 0.06 $ & -----  & ----- \\
IRS 16NW (IR-Paumard) & ----- & ----- & $ 199 \pm 52 $ & $ 67 \pm 44 $\\
\\
IRS 16SE2 (7mm) & $ 6.63 \pm 1.06 $ & $ 2.26 \pm 2.09 $ & $ 265.4 \pm 42.4 $ & $ 90.3 \pm 83.5 $\\
IRS 16SE2 (IR-Paumard) & -----  & ----- & $ 107 \pm 28 $ & $ 181 \pm 29 $\\
\\
IRS 16SW (7mm)  & $ 4.95 \pm 1.79 $ & $ 1.61 \pm 4.06 $ & $ 198.0 \pm 71.4 $ & $ 64.4 \pm 162.5 $\\
IRS 16SW (IR-Lu)  & $ 6.80 \pm 0.05 $ & $ 2.22 \pm 0.06 $ & ----- & ----- \\
IRS 16SW (IR-Paumard) & ----- & ----- & $ 261 \pm 47 $ & $ 90 \pm 43 $\\
\\
IRS 34W (7mm)  & $ -0.81 \pm 1.01 $ & $ -3.57 \pm 1.37 $ & $ -32.4 \pm 40.6 $ & $ -142.7 \pm 55.0 $\\
IRS 34W (IR-Paumard) & ----- & -----  & $ -79 \pm 28 $ & $ -166 \pm 27 $\\
\\
AFNW (7mm) & $ -3.89 \pm 0.95 $ & $ -0.32 \pm 1.37 $ & $ -155.8 \pm 38.1 $ & $ -12.9 \pm 54.7 $\\
AFNW (IR-Paumard) & ----- & ----- & $ -67 \pm 28 $ & $ -92 \pm 28 $\\
\\
AF (7mm)  & $ -2.40 \pm 0.83 $ & $ 8.14 \pm 1.31 $ & $ -96.1 \pm 33.2 $ & $ 325.5 \pm 52.4 $\\
AF (IR-Paumard) & -----  & ----- & $ 68 \pm 38 $ & $ 50 \pm 36 $\\
\\
SgrA* (7mm) & $ 0.000 \pm 0.000 $ & $ 0.000 \pm 0.000 $ & $ 0.0 \pm 0.0 $ & $ 0.0 \pm 0.0 $\\
\enddata
\end{deluxetable}


\begin{deluxetable}{lcccccc}
\tablecaption{Comparison of the absolute coordinates of the orbit-evolved 
NIR sources  from Lu \etal\,  (2009)  at the epoch of  radio data. 
Sgr A*'s  pixel position 3456.0$\times$3456.0 with the pixel size 
8.68159 mas.}
\tabletypesize{\scriptsize}
\tablecolumns{7}
\tablewidth{0pt}
\setlength{\tabcolsep}{0.04in}            
\tablehead{
\colhead{Name} & \colhead{dX($''$)} & \colhead{X$_{error}$} & \colhead{dY($''$)} &
\colhead{Y$_{error}$}  &\colhead{X } & \colhead{Y }\\
& (\sl{arcsec}) & (\sl{arcsec}) & (\sl{arcsec}) & (\sl{arcsec}) & (\sl{pixel number}) & (\sl{pixel number}) 
}
\startdata
IRS 16NW  & 0.1086 &  $\pm$0.0009 & 1.2352 &  $\pm$0.0033 & 3445.73 & 3596.75\\
IRS 16C   & 1.0078 &  $\pm$0.0023 &   0.5893 &  $\pm$0.0012 & 3336.02 & 3520.02\\
IRS 16SW  & 1.1326 &  $\pm$0.0025 &  -0.9334 &  $\pm$0.0023 & 3325.75 & 3346.70\\
IRS 16NE &  2.9041 &  $\pm$0.0034 &  0.9068 &  $\pm$0.0015 & 3121.91 & 3564.57\\
S3-5      & 2.9561  & $\pm$0.0049 & -1.1374 &  $\pm$0.0022 & 3115.41 & 3325.79\\
IRS 33E    & 0.7026  & $\pm$0.0054  & -3.1266  & $\pm$0.0107 & 3371.53 & 3093.82\\
\enddata
\end{deluxetable}

\begin{deluxetable}{lccccccccc}
\tablecaption{Comparison of the absolute coordinates of the orbit-evolved 
NIR sources  from  Paumard \etal\, (2006) at the epoch of  radio data. }
\tabletypesize{\scriptsize}
\tabletypesize{\scriptsize}
\tablecolumns{7}
\tablewidth{0pt}
\setlength{\tabcolsep}{0.04in}            
\tablehead{
\colhead{Source number} & {Name} & RA (IR)  & Dec (IR) & \colhead{dX} & \colhead{dY} &
   \colhead{X (NIR)} & \colhead{Y (NIR)} & \colhead{X (Radio)} & \colhead{Y (Radio)}\\
                        &        &  \sl{$17^ 45'$} & \sl{$-29^\circ 00'$} & \sl{arcsec} & \sl{arcsec} & 
\sl{pixel number} & \sl{pixel number} 
& \sl{pixel number} & \sl{pixel number} 
}
\startdata
E19 & IRS16NW &  40$^s$.0442 & $26'$.843  & 0.078 &  1.226 & 3447.015 & 3597.218& 3443.671 & 3595.712\\
E20 & IRS16C  &  40$^s$.1182 & 27$''$.517 &  1.048 &  0.552 &  3335.285&  3519.583&  3339.503& 3522.835\\
E23 & IRS16SW &  40$^s$.1231 & 29$''$.027 & 1.113 &  -0.958 & 3327.798 & 3345.652 &  3324.467 &  3347.939\\
E25 & W14     & 39$^s$.9182  & 28$''$.588 & -1.575 & -0.519 & 3637.418 & 3396.218 & 3639.259 & 3398.188\\
E39 & IRS16NE & 40$^s$.2590 & 27$''$.130  & 2.895 & 0.939 & 3122.536 & 3564.160 & 3121.024 & 3561.288\\
E40 & IRS16SE2 & 40$^s$.2644 & 29$''$.216 &  2.966 & -1.147 & 3114.358 & 3323.881 & 3113.503 & 3326.325\\
E41 & IRS33E   & 40$^s$.0912 & 31$''$.191  & 0.694 & -3.122 & 3376.061 & 3096.388 & 3371.388 & 3093.664\\
E51 & IRS13E2  & 39$^s$.7934 & 29$''$.793 & -3.213 & -1.724 & 3826.094 & 3257.419 & 3827.974 & 3257.590\\
E56 & IRS34W   & 39$^s$.7281 & 26$''$.519 & -4.069  & 1.550 & 3924.693 & 3634.539 & 3925.664 & 3631.678\\
E58 & IRS3E    & 39$^s$.8660 & 24$''$.269 & -2.260  & 3.800 & 3716.321 & 3893.708 & 3721.872 & 3895.693\\
E63 & IRS1W    & 40$^s$.4380 & 27$''$.449  & 5.244  & 0.620 & 2851.963 & 3527.415 & 2854.267 & 3526.253\\
E65 & IRS9W    & 40$^s$.2586 & 33$''$.657  & 2.890 & -5.588 & 3123.112 & 2812.339 & 3119.869 & 2812.905\\
E66 & IRS7SW   & 39$^s$.7371 & 23$''$.165 & -3.951  & 4.904 & 3911.101 & 4020.874 & 3910.743 & 4019.360\\             
E70 & IRS7E2   & 40$^s$.3782 & 23$''$.101 &  4.459  & 4.968 & 2942.384 & 4028.245 & 2939.391 & 4026.572\\             
E74 & AFNW     & 39$^s$.4555 & 31$''$.661 & -7.646 & -3.592 & 4336.714 & 3042.251 & 4334.515 & 3039.248\\             
E79 & AF       & 39$^s$.5410 & 34$''$.967 & -6.524 & -6.898 & 4207.475 & 2661.445 & 4203.622 & 2662.343\\             
E80 & IRS9SE   & 40$^s$.4690 & 36$''$.270  & 5.650 & -8.201 & 2805.198 & 2511.357 & 2805.641 & 2511.060\\             
E81 & AFNWNW   & 39$^s$.3058 & 30$''$.651 & -9.609 & -2.582 & 4562.825 & 3158.589 & 4560.840 & 3161.553\\             
E82 & Blum     & 39$^s$.3794 & 33$''$.339 & -8.643 & -5.270 & 4451.555 & 2848.968 & 4450.125 & 2848.228\\             
E83 & IRS15SW  & 39$^s$.9169 & 18$''$.057 & -1.593 & 10.012 & 3639.492 & 4609.245 & 3638.251 & 4609.362\\             
E88 & IRS15NE  & 40$^s$.1433 & 16$''$.364  & 1.378 & 11.705 & 3297.273 & 4804.255 & 3298.914 & 4803.245\\             
\enddata
\end{deluxetable}
 

\begin{deluxetable}{lcccccccc}           
\tablecaption{Derived mass loss rates from  radio and NIR observations}
\tabletypesize{\scriptsize}
\tablecolumns{6}
\tablewidth{0pt}
\setlength{\tabcolsep}{0.04in}
\tablehead{IR Name 
& \colhead{Velocity\tablenotemark{a}}
& \colhead{H/He}
& \colhead{Z}
& \colhead{$\dot{M}$ (NIR)} 
& \colhead{Flux\tablenotemark{b} (34.5 GHz)} 
& \colhead{$\dot{M}$ (34.5 GHz)}
&  \colhead{Flux\tablenotemark{b} (44.6 GHz)} 
& \colhead{$\dot{M}$ (44.6 GHz)} \\
& (km s$^{-1}$) 
&  
&   
& \sl{($10^{-5}$M$_{\odot}$ yr$^{-1}$)}  
& \sl{(mJy)}  
& \sl{($10^{-5}$M$_{\odot}$ yr$^{-1}$)}  
& \sl{(mJy)} 
& \sl{($10^{-5}$M$_{\odot}$ yr$^{-1}$)}  
}
\startdata
34W      & 650 &4.0 &1.33& 1.32 & 0.480$\pm$0.015  & 1.28$\pm0.03$                & 0.580$\pm0.038$ &  1.31$\pm0.02$ \\
IRS 16NW & 600 &5.0 &1.29& 1.12  & 0.228$\pm0.015$ & 0.66$\pm0.03$              & 0.245$\pm0.037$ &   0.62$\pm0.03$\\
IRS 16C  & 650 &2.5 &1.44& 2.24  & 0.449$\pm0.015$ & 1.26$\pm0.03$             & 0.525$\pm0.038$ &  1.266$\pm0.03$\\
33E      & 450 &0.26&1.33& 1.58 & 0.258$\pm0.014$ & 0.55$\pm0.02$             & 0.341$\pm0.038$ &  0.61$\pm0.02$\\
AF       & 700 &0.65&1.50& 1.78 & 0.650$\pm0.015$ & 1.83$\pm0.03$              & 0.767$\pm0.038$ &  1.85$\pm0.03$\\
AFNW     & 800 &0.39&1.95& 3.16  & 0.391$\pm0.015$& 1.63$\pm0.03$             & 0.476$\pm0.038$ &  1.68$\pm0.04$\\
IRS 16SE2 &2500&0.32&0.04& 7.08 & 0.325$\pm0.015$& 4.49$\pm0.15$              & 0.367$\pm0.038$ &  4.38$\pm0.13$\\
IRS 15NE  & 800 &0.26&2.0& 1.99 & 0.265 $\pm0.015$& 1.23$\pm0.05$              &0.370$\pm0.038$ & 1.41$\pm0.04$  \\
IRS 9W  & 1100 &0.1&1.95& 4.47 &0.201 $\pm0.014$    & 1.36$\pm0.07$              &0.331$\pm0.038$ &  1.76$\pm0.05$ \\ 
IRS 7SW  & 900 &0.0&2.0& 1.99 &0.146 $\pm0.015$    & 0.89$\pm0.07$              &0.143$\pm0.037$ & 0.78$\pm0.06$ \\
AFNWNW & 1800 &0.1&1.95& 1.22 & 0.234 $\pm0.015$   &2.50 $\pm0.01$              & 0.290$\pm0.038$&  2.61$\pm0.10$\\
\\
IRS 16NE\tablenotemark{c}  & 700 &-- & 0.98&-- & 0.933$\pm0.015$  &  1.94$\pm0.02$               &1.176$\pm0.038$ &  2.06$\pm0.02$\\
IRS 16SW\tablenotemark{c}  & 700 &--& 0.98&--  &0.242$\pm0.015$  &  0.70$\pm0.03$               &  0.297$\pm0.038$ &   0.73$\pm0.03$\\
\enddata
\tablenotetext{a}{Terminal wind velocity}
\tablenotetext{b}{Peak flux density}
\tablenotetext{c}{Terminal velocity is not available and  is assumed to be 700 \kms}
\end{deluxetable}


\begin{deluxetable}{llcccclcccccl}
\tablecaption{Gaussian Fitted Parameters of 34.5 GHz Continuum Sources within 30$''$ of Sgr~A*}
\label{Tab:Sources}
\rotate
\tabletypesize{\tiny}
\tablecolumns{12}
\tablewidth{0pt}
\setlength{\tabcolsep}{0.03in}
\tablehead{
\colhead{ID} & \colhead{Alt Name} & \colhead{RA (J2000)} & \colhead{Dec (J2000)} & \colhead{Dist. from Sgr A*} & \colhead{Pos. Precision} & \colhead{$ \theta_{a} \times \theta_{b} $ (PA)} & \colhead{Peak Intensity} & \colhead{Integrated Flux} &  \colhead{Flux Ks} & \colhead{Flux L'} &\colhead{Refs \& Comments} &\\
& & & & (\sl{arcsec}) & (\sl{mas}) & \sl{mas} $ \times $ \sl{mas} (\sl{deg}) & \sl{(mJy beam$^{-1}$)} & \sl{(mJy)} & \sl{(mJy)}& \sl{(mJy)}&
}
\startdata
1 & Sgr A* & 17:45:40.0383 & -29:00:28.0690 & 0.00 & 0.00 & - & 1719.700 $\pm$ 0.015 & 1722.900 $\pm$ 0.025 & 0.30 $\pm$ 0.016 & 1.47 $\pm$ 0.077 & - - \\
2 & - & 17:45:40.0072 & -29:00:28.0293 & 0.41 & 3.81 & 68 $\times$ 28(100) & 0.144 $\pm$ 0.014 & 0.266 $\pm$ 0.037 & - & 0.91 $\pm$ 0.048 & - a \\
3 & IRS 16C & 17:45:40.1152 & -29:00:27.4849 & 1.17 & 1.24 & 19 $\times$ 7(16) & 0.449 $\pm$ 0.015 & 0.465 $\pm$ 0.026 & 80.49 $\pm$ 4.003 & 138.05 $\pm$ 7.247 & 1 - \\
4 & IRS 16NW & 17:45:40.0463 & -29:00:26.8493 & 1.22 & 2.23 & 21 $\times$ 0(24) & 0.228 $\pm$ 0.015 & 0.215 $\pm$ 0.024 & 57.61 $\pm$ 2.865 & 97.29 $\pm$ 5.107 & 1 - \\
5 & n4 & 17:45:40.0076 & -29:00:29.2577 & 1.26 & 4.94 & 86 $\times$ 15(1) & 0.150 $\pm$ 0.014 & 0.220 $\pm$ 0.032 & 0.17 $\pm$ 0.010 & 1.18 $\pm$ 0.062 & - - \\
6 & - & 17:45:40.0216 & -29:00:29.4645 & 1.41 & 3.46 & 103 $\times$ 53(75) & 0.175 $\pm$ 0.014 & 0.494 $\pm$ 0.050 & - & 1.19 $\pm$ 0.062 & - a \\
7 & - & 17:45:39.9576 & -29:00:29.0822 & 1.47 & 7.25 & 103 $\times$ 42(162) & 0.107 $\pm$ 0.014 & 0.228 $\pm$ 0.041 & - & 0.50 $\pm$ 0.026 & - a \\
8 & IRS 16SW & 17:45:40.1255 & -29:00:29.0013 & 1.48 & 2.52 & 48 $\times$ 8(17) & 0.242 $\pm$ 0.014 & 0.287 $\pm$ 0.028 & 72.20 $\pm$ 3.591 & 116.32 $\pm$ 6.106 & 1 - \\
9 & - & 17:45:39.9237 & -29:00:28.4777 & 1.56 & 2.05 & - & 0.181  $\pm$  0.015 & 0.015  $\pm$  0.006 & - & - & - - \\
10 & - & 17:45:39.8934 & -29:00:28.1997 & 1.91 & 2.69 & 63 $\times$ 49(34) & 0.233 $\pm$ 0.014 & 0.428 $\pm$ 0.037 & - & 0.74 $\pm$ 0.039 & - a \\
11 & IRS 29 & 17:45:39.9214 & -29:00:26.7143 & 2.05 & 9.79 & 54 $\times$ 23(2) & 0.064 $\pm$ 0.014 & 0.084 $\pm$ 0.030 & 73.41 $\pm$ 3.651 & 596.01 $\pm$ 31.289 & 1 - \\
12 & IRS 16CC & 17:45:40.1891 & -29:00:27.4488 & 2.07 & 12.51 & 95 $\times$ 33(175) & 0.061 $\pm$ 0.014 & 0.110 $\pm$ 0.036 & 41.85 $\pm$ 2.082 & 60.60 $\pm$ 3.181 & 1 - \\
13 & Magnetar & 17:45:40.1680 & -29:00:29.7501 & 2.39 & 0.44 & 30 $\times$ 0(156) & 1.303  $\pm$  0.015 & 1.390  $\pm$  0.026 & - & - & - - \\
14 & - & 17:45:39.9099 & -29:00:29.7777 & 2.40 & 5.24 & 97 $\times$ 43(142) & 0.133  $\pm$  0.014 & 0.297  $\pm$  0.042 & - & - & - - \\
15 & - & 17:45:39.9108 & -29:00:26.3431 & 2.40 & 7.55 & 58 $\times$ 0(160) & 0.084  $\pm$  0.014 & 0.104  $\pm$  0.029 & - & - & - - \\
16 & - & 17:45:39.9312 & -29:00:30.2056 & 2.56 & 7.87 & 85 $\times$ 0(36) & 0.085  $\pm$  0.014 & 0.137  $\pm$  0.034 & - & - & - - \\
17 & n8 & 17:45:39.9088 & -29:00:30.3123 & 2.81 & 3.80 & 62 $\times$ 0(127) & 0.127 $\pm$ 0.015 & 0.001 $\pm$ 0.002 & 3.70 $\pm$ 0.187 & 6.24 $\pm$ 0.327 & - - \\
18 & - & 17:45:39.8329 & -29:00:28.9084 & 2.82 & 2.95 & 107 $\times$ 80(126) & 0.244  $\pm$  0.013 & 0.805  $\pm$  0.056 & - & - & - - \\
19 & - & 17:45:39.9456 & -29:00:30.7269 & 2.92 & 1.44 & 156 $\times$ 0(75) & 0.219  $\pm$  0.015 & 0.081  $\pm$  0.014 & - & - & - - \\
20 & - & 17:45:39.8379 & -29:00:29.4185 & 2.96 & 1.28 & 157 $\times$ 60(16) & 0.783  $\pm$  0.013 & 2.661  $\pm$  0.057 & - & - & - - \\
21 & IRS 13N-$\delta$ & 17:45:39.8163 & -29:00:28.8580 & 3.02 & 0.75 & 127 $\times$ 83(110) & 0.944 $\pm$ 0.013 & 3.726 $\pm$ 0.064 & 0.16 $\pm$ 0.010 & 12.36 $\pm$ 0.649 & 2 - \\
22 & IRS 16NE & 17:45:40.2600 & -29:00:27.1540 & 3.05 & 0.60 & 19 $\times$ 11(151) & 0.933 $\pm$ 0.015 & 0.987 $\pm$ 0.026 & 148.77 $\pm$ 7.399 & 266.96 $\pm$ 14.015 & 1 - \\
23 & IRS 13N-$\alpha$ & 17:45:39.8326 & -29:00:29.5086 & 3.06 & 1.24 & 117 $\times$ 77(139) & 0.617 $\pm$ 0.013 & 2.080  $\pm$  0.057 & 0.20 $\pm$ 0.011 & - & 2 b \\
24 & - & 17:45:39.8102 & -29:00:28.7058 & 3.06 & 3.45 & 116 $\times$ 42(152) & 0.230  $\pm$  0.014 & 0.550  $\pm$  0.044 & - & - & - - \\
25 & IRS 13N-$\beta$ & 17:45:39.8179 & -29:00:29.2564 & 3.13 & 1.31 & 113 $\times$ 70(164) & 0.614 $\pm$ 0.014 & 1.810 $\pm$ 0.051 & 0.30 $\pm$ 0.016 & 13.52 $\pm$ 0.710 & 2 - \\
26 & - & 17:45:39.7990 & -29:00:28.5348 & 3.17 & 2.73 & 138 $\times$ 75(138) & 0.300  $\pm$  0.013 & 1.132  $\pm$  0.062 & - & - & - - \\
27 & - & 17:45:39.8069 & -29:00:28.9758 & 3.17 & 1.66 & 104 $\times$ 66(28) & 0.454  $\pm$  0.014 & 1.264  $\pm$  0.049 & - & - & - - \\
28 & IRS 35 & 17:45:40.2648 & -29:00:29.1993 & 3.18 & 1.80 & 40 $\times$ 22(153) & 0.325 $\pm$ 0.014 & 0.404 $\pm$ 0.029 & 11.80 $\pm$ 0.587 & 27.55 $\pm$ 1.446 & 1 - \\
29 & - & 17:45:39.8210 & -29:00:29.4799 & 3.18 & 1.47 & 88 $\times$ 51(129) & 0.442 $\pm$ 0.014 & 1.030  $\pm$  0.043 & 0.12 $\pm$ 0.006 & - & - b \\
30 & - & 17:45:39.8165 & -29:00:29.4354 & 3.21 & 2.00 & 91 $\times$ 52(168) & 0.367  $\pm$  0.014 & 0.795  $\pm$  0.041 & - & - & - - \\
31 & IRS 33SE & 17:45:40.0941 & -29:00:31.2151 & 3.23 & 2.52 & 68 $\times$ 23(22) & 0.258 $\pm$ 0.014 & 0.382 $\pm$ 0.032 & 58.47 $\pm$ 2.908 & 96.75 $\pm$ 5.079 & 1 - \\
32 & IRS 33SW & 17:45:40.0148 & -29:00:31.2840 & 3.23 & 4.79 & 148 $\times$ 90(4) & 0.202 $\pm$ 0.013 & 0.854 $\pm$ 0.068 & 24.67 $\pm$ 1.227 & 40.41 $\pm$ 2.121 & 1 - \\
33 & - & 17:45:39.7922 & -29:00:28.3776 & 3.24 & 2.30 & 166 $\times$ 101(93) & 0.326  $\pm$  0.013 & 1.828  $\pm$  0.086 & - & - & - - \\
34 & - & 17:45:39.9079 & -29:00:30.8415 & 3.26 & 2.20 & 90 $\times$ 71(60) & 0.306  $\pm$  0.014 & 0.843  $\pm$  0.049 & - & - & - - \\
35 & IRS 13N-$\kappa$ & 17:45:39.7961 & -29:00:28.8174 & 3.26 & 2.56 & 120 $\times$ 27(115) & 0.235 $\pm$ 0.014 & 0.639 $\pm$ 0.049 & - & 7.94 $\pm$ 0.417 & 2 a \\
36 & - & 17:45:39.8056 & -29:00:29.2515 & 3.27 & 1.02 & 108 $\times$ 58(59) & 0.651 $\pm$ 0.014 & 1.891  $\pm$  0.051 & 0.55 $\pm$ 0.027 & - & - b \\
37 & IRS 13N-$\gamma$ & 17:45:39.7986 & -29:00:29.0012 & 3.28 & 1.36 & 100 $\times$ 82(176) & 0.561 $\pm$ 0.013 & 1.716  $\pm$  0.053 & 0.53 $\pm$ 0.030 & - & 2 b \\
38 & - & 17:45:39.8247 & -29:00:29.7828 & 3.28 & 0.57 & 125 $\times$ 93(25) & 1.480  $\pm$  0.013 & 5.828  $\pm$  0.064 & - & - & - - \\
39 & - & 17:45:39.7874 & -29:00:28.5065 & 3.32 & 1.44 & 77 $\times$ 66(127) & 0.455  $\pm$  0.014 & 1.078  $\pm$  0.044 & - & - & - - \\
40 & - & 17:45:39.7918 & -29:00:28.8128 & 3.32 & 2.40 & 60 $\times$ 0(180) & 0.218  $\pm$  0.014 & 0.354  $\pm$  0.034 & - & - & - - \\
41 & - & 17:45:40.1882 & -29:00:30.7867 & 3.35 & 6.55 & 122 $\times$ 109(26) & 0.128  $\pm$  0.013 & 0.558  $\pm$  0.070 & - & - & - - \\
42 & - & 17:45:39.7993 & -29:00:29.2826 & 3.36 & 1.45 & 63 $\times$ 55(119) & 0.424  $\pm$  0.014 & 0.832  $\pm$  0.038 & - & - & - - \\
43 & - & 17:45:39.7837 & -29:00:28.3883 & 3.36 & 2.25 & 99 $\times$ 67(166) & 0.338  $\pm$  0.014 & 0.893  $\pm$  0.048 & - & - & - - \\
44 & - & 17:45:40.0964 & -29:00:31.3526 & 3.37 & 3.20 & 26 $\times$ 0(172) & 0.165  $\pm$  0.015 & 0.182  $\pm$  0.027 & - & - & - - \\
45 & - & 17:45:39.8474 & -29:00:30.4173 & 3.43 & 3.05 & 147 $\times$ 72(76) & 0.219 $\pm$ 0.013 & 0.933  $\pm$  0.068 & 0.09 $\pm$ 0.004 & - & - b \\
46 & - & 17:45:39.8212 & -29:00:29.9818 & 3.43 & 0.35 & - & 0.542  $\pm$  0.015 & 0.071  $\pm$  0.008 & - & - & - - \\
47 & - & 17:45:39.7868 & -29:00:29.0695 & 3.45 & 2.97 & 69 $\times$ 53(179) & 0.223 $\pm$ 0.014 & 0.430  $\pm$  0.038 & 0.05 $\pm$ 0.003 & - & - b \\
48 & - & 17:45:40.2093 & -29:00:30.7183 & 3.47 & 0.72 & 139 $\times$ 89(83) & 0.991  $\pm$  0.013 & 4.441  $\pm$  0.071 & - & - & - - \\
49 & - & 17:45:39.8095 & -29:00:29.8291 & 3.48 & 0.91 & 116 $\times$ 84(21) & 0.892  $\pm$  0.013 & 3.079  $\pm$  0.058 & - & - & - - \\
50 & - & 17:45:40.2193 & -29:00:30.6437 & 3.50 & 0.66 & - & 0.553  $\pm$  0.015 & 0.043  $\pm$  0.006 & - & - & - - \\
51 & - & 17:45:39.9345 & -29:00:31.2941 & 3.50 & 2.06 & 144 $\times$ 99(4) & 0.461 $\pm$ 0.013 & 2.081 $\pm$ 0.072 & - & 1.64 $\pm$ 0.086 & - a \\
52 & IRS 21 & 17:45:40.2175 & -29:00:30.7427 & 3.56 & 0.04 & 12 $\times$ 0(89) & 0.805 $\pm$ 0.015 & 0.044  $\pm$  0.005 & 41.51 $\pm$ 2.064 & - & 1 b \\
53 & IRS 13-E3c & 17:45:39.7933 & -29:00:29.6178 & 3.57 & 0.16 & 112 $\times$ 89(37) & 4.913 $\pm$ 0.013 & 17.447 $\pm$ 0.059 & 5.36 $\pm$ 0.336 & 233.37 $\pm$ 12.251 & 2 - \\
54 & IRS 13-E3N & 17:45:39.7878 & -29:00:29.4705 & 3.57 & 0.66 & 89 $\times$ 48(128) & 0.975 $\pm$ 0.014 & 2.241 $\pm$ 0.043 & 14.42 $\pm$ 0.730 & 97.47 $\pm$ 5.117 & 2 - \\
55 & - & 17:45:40.2121 & -29:00:30.8404 & 3.59 & 0.95 & 108 $\times$ 71(140) & 0.777  $\pm$  0.013 & 2.353  $\pm$  0.052 & - & - & - - \\
56 & - & 17:45:40.2233 & -29:00:30.7224 & 3.60 & 1.17 & 144 $\times$ 114(10) & 0.806  $\pm$  0.013 & 4.080  $\pm$  0.079 & - & - & - - \\
57 & IRS 21 & 17:45:40.2171 & -29:00:30.8244 & 3.62 & 0.01 & - & 0.704 $\pm$ 0.015 & 0.002 $\pm$ 0.001 & - & 496.19 $\pm$ 26.049 & 1 a \\
58 & n17 & 17:45:39.9929 & -29:00:31.6362 & 3.62 & 3.27 & 97 $\times$ 86(170) & 0.229 $\pm$ 0.013 & 0.712 $\pm$ 0.054 & 0.15 $\pm$ 0.009 & 1.37 $\pm$ 0.072 & - - \\
59 & - & 17:45:40.2063 & -29:00:30.9615 & 3.64 & 9.09 & 155 $\times$ 72(10) & 0.110  $\pm$  0.013 & 0.411  $\pm$  0.062 & - & - & - - \\
60 & - & 17:45:40.2200 & -29:00:30.8343 & 3.65 & 1.65 & 166 $\times$ 126(49) & 0.574  $\pm$  0.013 & 3.628  $\pm$  0.095 & - & - & - - \\
61 & IRS 13-E5 & 17:45:39.7776 & -29:00:29.5497 & 3.73 & 0.48 & 92 $\times$ 60(23) & 1.505 $\pm$ 0.014 & 3.654 $\pm$ 0.045 & 1.15 $\pm$ 0.060 & 41.69 $\pm$ 2.189 & 2 - \\
62 & - & 17:45:39.7683 & -29:00:29.4468 & 3.80 & 2.14 & 68 $\times$ 54(173) & 0.306  $\pm$  0.014 & 0.594  $\pm$  0.038 & - & - & - - \\
63 & n19 & 17:45:40.3112 & -29:00:26.5384 & 3.89 & 3.39 & 95 $\times$ 84(99) & 0.206 $\pm$ 0.013 & 0.646 $\pm$ 0.054 & 0.09 $\pm$ 0.006 & 4.40 $\pm$ 0.231 & - - \\
64 & - & 17:45:39.7944 & -29:00:30.4545 & 3.99 & 1.72 & 325 $\times$ 91(3) & 1.084 $\pm$ 0.013 & 9.082  $\pm$  0.121 & 0.09 $\pm$ 0.005 & - & - b \\
65 & - & 17:45:39.7597 & -29:00:29.7418 & 4.02 & 1.23 & 112 $\times$ 49(116) & 0.516  $\pm$  0.014 & 1.474  $\pm$  0.050 & - & - & - - \\
66 & n20 & 17:45:40.3272 & -29:00:26.4595 & 4.12 & 4.35 & 97 $\times$ 90(79) & 0.165 $\pm$ 0.013 & 0.546 $\pm$ 0.056 & 6.32 $\pm$ 0.314 & 8.20 $\pm$ 0.430 & - - \\
67 & - & 17:45:39.7451 & -29:00:26.5452 & 4.14 & 4.26 & 213 $\times$ 58(146) & 0.263 $\pm$ 0.013 & 1.226  $\pm$  0.073 & 0.15 $\pm$ 0.009 & - & - b \\
68 & - & 17:45:39.7664 & -29:00:25.8938 & 4.18 & 3.46 & 271 $\times$ 173(150) & 0.379 $\pm$ 0.012 & 4.738  $\pm$  0.156 & 0.04 $\pm$ 0.002 & - & - b \\
69 & - & 17:45:39.7515 & -29:00:29.9388 & 4.20 & 1.37 & - & 0.273  $\pm$  0.015 & 0.023  $\pm$  0.006 & - & - & - - \\
70 & - & 17:45:39.7562 & -29:00:30.1631 & 4.25 & 1.75 & - & 0.273 $\pm$ 0.015 & 0.009  $\pm$  0.004 & 0.05 $\pm$ 0.003 & - & - b \\
71 & IRS 34W & 17:45:39.7275 & -29:00:26.5423 & 4.35 & 1.11 & 31 $\times$ 0(26) & 0.481 $\pm$ 0.015 & 0.526 $\pm$ 0.027 & 21.39 $\pm$ 1.064 & 49.16 $\pm$ 2.581 & 3 - \\
72 & IRS 3 & 17:45:39.8620 & -29:00:24.2480 & 4.47 & 9.71 & 94 $\times$ 0(148) & 0.074 $\pm$ 0.014 & 0.093 $\pm$ 0.029 & 20.84 $\pm$ 1.037 & 1252.14 $\pm$ 65.733 & 4 - \\
73 & - & 17:45:39.7488 & -29:00:30.6291 & 4.58 & 2.20 & 173 $\times$ 45(14) & 0.490 $\pm$ 0.013 & 1.546  $\pm$  0.054 & 0.05 $\pm$ 0.002 & - & - b \\
74 & - & 17:45:40.2378 & -29:00:31.8304 & 4.58 & 7.54 & 72 $\times$ 49(59) & 0.081  $\pm$  0.014 & 0.167  $\pm$  0.040 & - & - & - - \\
75 & - & 17:45:40.3273 & -29:00:25.3226 & 4.68 & 12.38 & 305 $\times$ 0(104) & 0.039  $\pm$  0.015 & 0.001  $\pm$  0.003 & - & - & - - \\
76 & - & 17:45:39.7299 & -29:00:30.5161 & 4.73 & 3.14 & 70 $\times$ 53(150) & 0.206  $\pm$  0.014 & 0.408  $\pm$  0.039 & - & - & - - \\
77 & IRS 34NW & 17:45:39.7494 & -29:00:25.2459 & 4.73 & 18.67 & 53 $\times$ 13(134) & 0.031 $\pm$ 0.014 & 0.043 $\pm$ 0.031 & 3.65 $\pm$ 0.185 & 6.05 $\pm$ 0.318 & 3 - \\
78 & - & 17:45:40.2622 & -29:00:31.7715 & 4.73 & 8.94 & 89 $\times$ 41(170) & 0.082  $\pm$  0.014 & 0.157  $\pm$  0.038 & - & - & - - \\
79 & - & 17:45:39.7494 & -29:00:25.2458 & 4.73 & 18.54 & 54 $\times$ 14(131) & 0.031  $\pm$  0.014 & 0.044  $\pm$  0.031 & - & - & - - \\
80 & - & 17:45:39.7422 & -29:00:31.1574 & 4.96 & 1.26 & 140 $\times$ 109(149) & 0.710 $\pm$ 0.013 & 3.432  $\pm$  0.076 & 0.06 $\pm$ 0.003 & - & - b \\
81 & - & 17:45:39.7276 & -29:00:25.2467 & 4.96 & 8.98 & 102 $\times$ 28(145) & 0.080  $\pm$  0.014 & 0.163  $\pm$  0.040 & - & - & - - \\
82 & - & 17:45:39.7275 & -29:00:25.2450 & 4.96 & 10.13 & 125 $\times$ 56(148) & 0.080  $\pm$  0.014 & 0.230  $\pm$  0.051 & - & - & - - \\
83 & - & 17:45:40.2748 & -29:00:32.0907 & 5.08 & 17.69 & 168 $\times$ 43(159) & 0.058  $\pm$  0.013 & 0.180  $\pm$  0.053 & - & - & - - \\
84 & - & 17:45:39.6380 & -29:00:28.1141 & 5.25 & 2.93 & 243 $\times$ 83(73) & 0.261 $\pm$ 0.013 & 1.878 $\pm$ 0.106 & - & 0.96 $\pm$ 0.050 & - a \\
85 & IRS 1W & 17:45:40.4362 & -29:00:27.4499 & 5.26 & 0.53 & - & 0.443 $\pm$ 0.015 & 0.083 $\pm$ 0.009 & 80.86 $\pm$ 4.021 & 554.69 $\pm$ 29.120 & 4 - \\
86 & - & 17:45:39.6831 & -29:00:25.5558 & 5.29 & 7.25 & 182 $\times$ 65(153) & 0.146  $\pm$  0.013 & 0.603  $\pm$  0.067 & - & - & - - \\
87 & - & 17:45:39.6561 & -29:00:26.0476 & 5.41 & 5.57 & 95 $\times$ 17(156) & 0.132  $\pm$  0.014 & 0.224  $\pm$  0.035 & - & - & - - \\
88 & - & 17:45:39.6226 & -29:00:26.5877 & 5.65 & 1.02 & 118 $\times$ 76(114) & 0.678 $\pm$ 0.013 & 2.390 $\pm$ 0.059 & - & 1.89 $\pm$ 0.099 & - a \\
89 & n25 & 17:45:40.1050 & -29:00:22.2313 & 5.90 & 4.36 & 39 $\times$ 0(151) & 0.134 $\pm$ 0.015 & 0.146 $\pm$ 0.027 & 13.32 $\pm$ 0.662 & 23.97 $\pm$ 1.258 & - - \\
90 & - & 17:45:39.5712 & -29:00:29.0089 & 6.20 & 8.30 & 588 $\times$ 163(10) & 0.249  $\pm$  0.008 & 6.113  $\pm$  0.212 & - & - & - - \\
91 & IRS 9W & 17:45:40.2609 & -29:00:33.6449 & 6.29 & 2.98 & 44 $\times$ 16(19) & 0.201 $\pm$ 0.014 & 0.244 $\pm$ 0.028 & 10.16 $\pm$ 0.515 & 19.50 $\pm$ 1.024 & 3 - \\
92 & IRS 7SW & 17:45:39.7371 & -29:00:23.1803 & 6.29 & 3.41 & - & 0.146 $\pm$ 0.015 & 0.126 $\pm$ 0.023 & 7.40 $\pm$ 0.368 & 12.16 $\pm$ 0.638 & 3 - \\
93 & - & 17:45:39.6189 & -29:00:31.5673 & 6.52 & 8.89 & 105 $\times$ 58(24) & 0.086 $\pm$ 0.014 & 0.221 $\pm$ 0.047 & - & 0.11 $\pm$ 0.006 & - a \\
94 & - & 17:45:39.5481 & -29:00:26.3898 & 6.65 & 4.71 & 260 $\times$ 20(3) & 0.331  $\pm$  0.013 & 1.127  $\pm$  0.057 & - & - & - - \\
95 & - & 17:45:39.5658 & -29:00:25.6191 & 6.67 & 9.35 & 234 $\times$ 105(8) & 0.147  $\pm$  0.013 & 1.035  $\pm$  0.104 & - & - & - - \\
96 & n28 & 17:45:40.1550 & -29:00:21.5389 & 6.71 & 10.88 & 63 $\times$ 0(33) & 0.057 $\pm$ 0.014 & 0.069 $\pm$ 0.028 & 1.33 $\pm$ 0.070 & 3.99 $\pm$ 0.209 & - - \\
97 & - & 17:45:39.6574 & -29:00:23.3315 & 6.89 & 7.62 & 34 $\times$ 0(17) & 0.068  $\pm$  0.015 & 0.076  $\pm$  0.027 & - & - & - - \\
98 & - & 17:45:39.5534 & -29:00:24.9464 & 7.09 & 5.10 & 215 $\times$ 201(102) & 0.220  $\pm$  0.012 & 2.584  $\pm$  0.152 & - & - & - - \\
99 & - & 17:45:39.5276 & -29:00:25.5663 & 7.15 & 5.70 & 319 $\times$ 149(7) & 0.285  $\pm$  0.012 & 3.578  $\pm$  0.157 & - & - & - - \\
100 & - & 17:45:40.5213 & -29:00:24.0860 & 7.48 & 16.01 & 120 $\times$ 25(38) & 0.047  $\pm$  0.014 & 0.111  $\pm$  0.044 & - & - & - - \\
101 & - & 17:45:39.4794 & -29:00:26.4570 & 7.51 & 3.71 & 158 $\times$ 87(2) & 0.275  $\pm$  0.013 & 1.190  $\pm$  0.069 & - & - & - - \\
102 & - & 17:45:40.2035 & -29:00:20.7599 & 7.62 & 7.72 & 54 $\times$ 0(44) & 0.073  $\pm$  0.015 & 0.072  $\pm$  0.025 & - & - & - - \\
103 & - & 17:45:39.4983 & -29:00:24.9954 & 7.72 & 7.88 & 217 $\times$ 191(157) & 0.153  $\pm$  0.012 & 1.711  $\pm$  0.149 & - & - & - - \\
104 & - & 17:45:39.4390 & -29:00:28.1584 & 7.86 & 11.91 & 330 $\times$ 117(15) & 0.150 $\pm$ 0.013 & 1.576  $\pm$  0.145 & 0.03 $\pm$ 0.001 & - & - b \\
105 & - & 17:45:39.5412 & -29:00:23.5811 & 7.92 & 23.60 & 289 $\times$ 127(144) & 0.060  $\pm$  0.013 & 0.618  $\pm$  0.144 & - & - & - - \\
106 & - & 17:45:40.5607 & -29:00:32.0865 & 7.94 & 7.41 & 131 $\times$ 23(45) & 0.100  $\pm$  0.014 & 0.265  $\pm$  0.047 & - & - & - - \\
107 & - & 17:45:39.4510 & -29:00:25.9187 & 8.00 & 10.70 & 106 $\times$ 63(36) & 0.069 $\pm$ 0.014 & 0.192 $\pm$ 0.049 & - & 1.94 $\pm$ 0.102 & - a \\
108 & - & 17:45:40.5187 & -29:00:33.0163 & 8.01 & 10.47 & 79 $\times$ 59(160) & 0.065  $\pm$  0.014 & 0.143  $\pm$  0.042 & - & - & - - \\
109 & - & 17:45:40.4671 & -29:00:22.1546 & 8.16 & 5.85 & 89 $\times$ 73(128) & 0.117  $\pm$  0.014 & 0.317  $\pm$  0.049 & - & - & - - \\
110 & AFNW & 17:45:39.4569 & -29:00:31.6843 & 8.44 & 1.38 & 9 $\times$ 0(19) & 0.391 $\pm$ 0.015 & 0.387 $\pm$ 0.025 & 15.18 $\pm$ 0.755 & 24.78 $\pm$ 1.301 & 3 - \\
111 & - & 17:45:40.5908 & -29:00:23.5846 & 8.52 & 17.04 & 71 $\times$ 43(46) & 0.037  $\pm$  0.014 & 0.070  $\pm$  0.038 & - & - & - - \\
112 & - & 17:45:39.4848 & -29:00:32.7037 & 8.61 & 26.17 & 151 $\times$ 93(114) & 0.029  $\pm$  0.013 & 0.143  $\pm$  0.076 & - & - & - - \\
113 & n30 & 17:45:39.9455 & -29:00:36.6367 & 8.65 & 13.84 & 171 $\times$ 41(141) & 0.066 $\pm$ 0.013 & 0.228 $\pm$ 0.058 & 0.46 $\pm$ 0.025 & 0.54 $\pm$ 0.029 & - - \\
114 & - & 17:45:39.9360 & -29:00:36.6238 & 8.66 & 7.81 & 40 $\times$ 0(9) & 0.063  $\pm$  0.014 & 0.075  $\pm$  0.028 & - & - & - - \\
115 & - & 17:45:39.9395 & -29:00:36.6614 & 8.69 & 10.48 & 137 $\times$ 21(110) & 0.056  $\pm$  0.013 & 0.172  $\pm$  0.053 & - & - & - - \\
116 & - & 17:45:39.3677 & -29:00:28.0994 & 8.80 & 10.53 & 185 $\times$ 41(154) & 0.102  $\pm$  0.013 & 0.345  $\pm$  0.057 & - & - & - - \\
117 & - & 17:45:39.9039 & -29:00:36.7861 & 8.89 & 14.84 & 98 $\times$ 81(2) & 0.051  $\pm$  0.014 & 0.153  $\pm$  0.052 & - & - & - - \\
118 & - & 17:45:40.4043 & -29:00:20.5582 & 8.91 & 4.82 & 95 $\times$ 67(132) & 0.143  $\pm$  0.014 & 0.389  $\pm$  0.048 & - & - & - - \\
119 & - & 17:45:40.6562 & -29:00:24.0843 & 9.03 & 18.75 & 97 $\times$ 56(43) & 0.037  $\pm$  0.014 & 0.093  $\pm$  0.046 & - & - & - - \\
120 & - & 17:45:39.6671 & -29:00:20.4283 & 9.06 & 22.21 & 259 $\times$ 132(24) & 0.064  $\pm$  0.013 & 0.610  $\pm$  0.136 & - & - & - - \\
121 & - & 17:45:39.8385 & -29:00:19.2736 & 9.18 & 4.56 & 56 $\times$ 45(12) & 0.136  $\pm$  0.014 & 0.224  $\pm$  0.034 & - & - & - - \\
122 & - & 17:45:39.3419 & -29:00:29.0936 & 9.19 & 12.65 & 132 $\times$ 57(12) & 0.071  $\pm$  0.014 & 0.206  $\pm$  0.051 & - & - & - - \\
123 & - & 17:45:40.4140 & -29:00:20.1767 & 9.30 & 9.31 & 233 $\times$ 142(138) & 0.128 $\pm$ 0.013 & 1.191 $\pm$ 0.133 & - & 0.09 $\pm$ 0.005 & - a \\
124 & n31 & 17:45:40.7579 & -29:00:27.8216 & 9.44 & 6.72 & 57 $\times$ 0(130) & 0.084 $\pm$ 0.014 & 0.102 $\pm$ 0.028 & 4.85 $\pm$ 0.246 & 6.40 $\pm$ 0.336 & - - \\
125 & - & 17:45:40.4068 & -29:00:19.9339 & 9.46 & 2.56 & 111 $\times$ 38(130) & 0.264  $\pm$  0.014 & 0.665  $\pm$  0.046 & - & - & - - \\
126 & AF/AHH & 17:45:39.5435 & -29:00:34.9596 & 9.47 & 0.86 & 21 $\times$ 13(1) & 0.650 $\pm$ 0.015 & 0.694 $\pm$ 0.026 & 31.84 $\pm$ 1.583 & 58.35 $\pm$ 3.063 & 3 - \\
127 & - & 17:45:39.4133 & -29:00:32.8689 & 9.50 & 5.15 & 63 $\times$ 0(174) & 0.097  $\pm$  0.014 & 0.153  $\pm$  0.033 & - & - & - - \\
128 & - & 17:45:40.5946 & -29:00:21.9641 & 9.51 & 17.78 & 116 $\times$ 61(7) & 0.047  $\pm$  0.014 & 0.128  $\pm$  0.049 & - & - & - - \\
129 & - & 17:45:39.4472 & -29:00:33.6588 & 9.56 & 13.43 & 33 $\times$ 0(129) & 0.040 $\pm$ 0.015 & 0.037 $\pm$ 0.024 & - & 0.10 $\pm$ 0.005 & - - \\
130 & - & 17:45:39.5727 & -29:00:20.6684 & 9.60 & 25.23 & 430 $\times$ 107(21) & 0.080  $\pm$  0.012 & 1.022  $\pm$  0.158 & - & - & - - \\
131 & - & 17:45:40.6861 & -29:00:23.5839 & 9.61 & 4.29 & 78 $\times$ 30(101) & 0.129  $\pm$  0.014 & 0.262  $\pm$  0.039 & - & - & - - \\
132 & - & 17:45:39.5760 & -29:00:20.5667 & 9.65 & 14.58 & 192 $\times$ 69(15) & 0.080  $\pm$  0.013 & 0.350  $\pm$  0.070 & - & - & - - \\
133 & n33 & 17:45:39.8331 & -29:00:18.8059 & 9.65 & 26.38 & 230 $\times$ 71(138) & 0.042 $\pm$ 0.013 & 0.238 $\pm$ 0.087 & 0.35 $\pm$ 0.019 & 0.21 $\pm$ 0.013 & - - \\
134 & - & 17:45:40.4578 & -29:00:20.1335 & 9.66 & 9.40 & 81 $\times$ 39(171) & 0.075  $\pm$  0.014 & 0.133  $\pm$  0.036 & - & - & - - \\
135 & - & 17:45:40.4182 & -29:00:19.5835 & 9.84 & 8.74 & 207 $\times$ 126(43) & 0.123  $\pm$  0.013 & 0.944  $\pm$  0.112 & - & - & - - \\
136 & - & 17:45:40.4571 & -29:00:19.8616 & 9.88 & 8.51 & 85 $\times$ 57(54) & 0.076  $\pm$  0.014 & 0.182  $\pm$  0.044 & - & - & - - \\
137 & - & 17:45:39.7029 & -29:00:19.1808 & 9.92 & 13.77 & 146 $\times$ 125(27) & 0.068  $\pm$  0.013 & 0.378  $\pm$  0.085 & - & - & - - \\
138 & AFNWNW & 17:45:39.3070 & -29:00:30.6316 & 9.93 & 2.35 & 17 $\times$ 0(39) & 0.234 $\pm$ 0.015 & 0.238 $\pm$ 0.026 & 8.26 $\pm$ 0.418 & 15.68 $\pm$ 0.823 & 3 - \\
139 & - & 17:45:40.6251 & -29:00:21.7848 & 9.94 & 25.12 & 76 $\times$ 65(119) & 0.026  $\pm$  0.014 & 0.060  $\pm$  0.043 & - & - & - - \\
140 & - & 17:45:39.2976 & -29:00:30.3461 & 9.98 & 36.95 & 242 $\times$ 95(149) & 0.035  $\pm$  0.013 & 0.238  $\pm$  0.102 & - & - & - - \\
141 & - & 17:45:40.4351 & -29:00:19.5028 & 10.02 & 4.01 & 189 $\times$ 124(86) & 0.211 $\pm$ 0.013 & 1.522 $\pm$ 0.106 & - & 0.12 $\pm$ 0.007 & - a \\
142 & - & 17:45:39.2749 & -29:00:27.2890 & 10.04 & 1.71 & 81 $\times$ 58(173) & 0.409 $\pm$ 0.014 & 0.884  $\pm$  0.041 & 0.05 $\pm$ 0.003 & - & - b \\
143 & - & 17:45:39.3973 & -29:00:33.6179 & 10.07 & 26.78 & 347 $\times$ 208(16) & 0.053  $\pm$  0.010 & 0.981  $\pm$  0.186 & - & - & - - \\
144 & - & 17:45:39.2689 & -29:00:28.0515 & 10.09 & 5.28 & 121 $\times$ 59(26) & 0.154  $\pm$  0.014 & 0.445  $\pm$  0.051 & - & - & - - \\
145 & Blum & 17:45:39.3800 & -29:00:33.3474 & 10.12 & 8.64 & 22 $\times$ 0(151) & 0.064 $\pm$ 0.015 & 0.057 $\pm$ 0.023 & 4.59 $\pm$ 0.232 & 6.84 $\pm$ 0.359 & 3 - \\
146 & IRS 15SW & 17:45:39.9178 & -29:00:18.0503 & 10.14 & 3.40 & - & 0.142 $\pm$ 0.015 & 0.118 $\pm$ 0.022 & 12.15 $\pm$ 0.615 & 15.76 $\pm$ 0.828 & 3 - \\
147 & - & 17:45:40.1416 & -29:00:18.0117 & 10.15 & 220.18 & 9950 $\times$ 0(50) & 0.041  $\pm$  0.003 & 6.251  $\pm$  0.511 & - & - & - - \\
148 & - & 17:45:40.3945 & -29:00:18.7981 & 10.38 & 4.78 & 122 $\times$ 40(137) & 0.154 $\pm$ 0.014 & 0.408 $\pm$ 0.047 & - & 0.10 $\pm$ 0.005 & - a \\
149 & - & 17:45:39.2453 & -29:00:27.4434 & 10.42 & 3.20 & 245 $\times$ 189(79) & 0.324  $\pm$  0.012 & 4.085  $\pm$  0.157 & - & - & - - \\
150 & - & 17:45:39.2708 & -29:00:25.0090 & 10.52 & 4.89 & 62 $\times$ 21(72) & 0.112  $\pm$  0.014 & 0.190  $\pm$  0.035 & - & - & - - \\
151 & - & 17:45:39.2262 & -29:00:28.2596 & 10.66 & 7.55 & 10330 $\times$ 0(102) & 0.425  $\pm$  0.003 & 59.290  $\pm$  0.489 & - & - & - - \\
152 & - & 17:45:39.2635 & -29:00:24.8139 & 10.67 & 7.97 & 181 $\times$ 85(177) & 0.142 $\pm$ 0.013 & 0.673 $\pm$ 0.075 & - & 0.11 $\pm$ 0.006 & - a \\
153 & - & 17:45:40.4368 & -29:00:18.5960 & 10.82 & 4.21 & 122 $\times$ 14(118) & 0.145  $\pm$  0.014 & 0.380  $\pm$  0.047 & - & - & - - \\
154 & - & 17:45:39.9449 & -29:00:17.3037 & 10.83 & 13.58 & 133 $\times$ 119(4) & 0.066  $\pm$  0.013 & 0.325  $\pm$  0.077 & - & - & - - \\
155 & - & 17:45:39.2117 & -29:00:27.6703 & 10.85 & 32.75 & 957 $\times$ 61(7) & 0.120  $\pm$  0.010 & 2.163  $\pm$  0.184 & - & - & - - \\
156 & - & 17:45:39.8886 & -29:00:17.2578 & 10.99 & 20.35 & 159 $\times$ 114(102) & 0.040  $\pm$  0.013 & 0.234  $\pm$  0.089 & - & - & - - \\
157 & - & 17:45:39.2409 & -29:00:23.7826 & 11.30 & 14.83 & 188 $\times$ 126(170) & 0.077  $\pm$  0.013 & 0.525  $\pm$  0.101 & - & - & - - \\
158 & - & 17:45:40.7220 & -29:00:21.0700 & 11.38 & 0.20 & 120 $\times$ 0(89) & 0.098  $\pm$  0.015 & 0.004  $\pm$  0.004 & - & - & - - \\
159 & - & 17:45:39.1603 & -29:00:28.1764 & 11.52 & 6.51 & 235 $\times$ 144(81) & 0.145  $\pm$  0.013 & 1.427  $\pm$  0.140 & - & - & - - \\
160 & - & 17:45:39.2284 & -29:00:23.5970 & 11.53 & 8.66 & 161 $\times$ 44(26) & 0.113  $\pm$  0.013 & 0.353  $\pm$  0.054 & - & - & - - \\
161 & - & 17:45:39.1555 & -29:00:27.8122 & 11.58 & 15.82 & 289 $\times$ 79(13) & 0.104  $\pm$  0.013 & 0.712  $\pm$  0.102 & - & - & - - \\
162 & IRS 15NE & 17:45:40.1423 & -29:00:16.3713 & 11.78 & 2.14 & 31 $\times$ 0(145) & 0.265 $\pm$ 0.015 & 0.285 $\pm$ 0.026 & 14.10 $\pm$ 0.714 & 23.06 $\pm$ 1.211 & 3 - \\
163 & - & 17:45:39.1649 & -29:00:25.2529 & 11.80 & 12.45 & 170 $\times$ 46(6) & 0.087  $\pm$  0.013 & 0.269  $\pm$  0.053 & - & - & - - \\
164 & n38 & 17:45:40.9554 & -29:00:27.8892 & 12.03 & 14.40 & 232 $\times$ 48(27) & 0.088 $\pm$ 0.013 & 0.395 $\pm$ 0.071 & 0.15 $\pm$ 0.008 & 0.17 $\pm$ 0.011 & - - \\
165 & - & 17:45:40.1836 & -29:00:16.1777 & 12.04 & 12.79 & 121 $\times$ 106(85) & 0.061 $\pm$ 0.013 & 0.265 $\pm$ 0.070 & - & 0.07 $\pm$ 0.004 & - a \\
166 & - & 17:45:39.1207 & -29:00:27.7423 & 12.04 & 16.11 & 268 $\times$ 121(30) & 0.088  $\pm$  0.013 & 0.806  $\pm$  0.132 & - & - & - - \\
167 & - & 17:45:39.1206 & -29:00:27.7443 & 12.04 & 15.51 & 252 $\times$ 119(28) & 0.088  $\pm$  0.013 & 0.746  $\pm$  0.123 & - & - & - - \\
168 & - & 17:45:40.4443 & -29:00:17.2614 & 12.05 & 12.66 & 186 $\times$ 141(90) & 0.073  $\pm$  0.013 & 0.563  $\pm$  0.113 & - & - & - - \\
169 & - & 17:45:40.4442 & -29:00:17.2611 & 12.05 & 12.27 & 158 $\times$ 133(78) & 0.073  $\pm$  0.013 & 0.463  $\pm$  0.096 & - & - & - - \\
170 & - & 17:45:39.1166 & -29:00:29.4422 & 12.17 & 22.24 & 117 $\times$ 19(41) & 0.033  $\pm$  0.014 & 0.075  $\pm$  0.043 & - & - & - - \\
171 & - & 17:45:40.8327 & -29:00:34.3871 & 12.19 & 5.79 & 226 $\times$ 90(44) & 0.188  $\pm$  0.013 & 1.229  $\pm$  0.098 & - & - & - - \\
172 & - & 17:45:40.9771 & -29:00:28.0832 & 12.31 & 15.58 & 259 $\times$ 227(14) & 0.078  $\pm$  0.011 & 1.197  $\pm$  0.171 & - & - & - - \\
173 & - & 17:45:40.4553 & -29:00:16.9167 & 12.42 & 9.37 & 211 $\times$ 135(10) & 0.134  $\pm$  0.013 & 1.064  $\pm$  0.116 & - & - & - - \\
174 & - & 17:45:39.8825 & -29:00:15.7784 & 12.46 & 24.58 & 72 $\times$ 30(179) & 0.028  $\pm$  0.014 & 0.042  $\pm$  0.033 & - & - & - - \\
175 & - & 17:45:39.8825 & -29:00:15.7810 & 12.46 & 25.06 & 74 $\times$ 22(2) & 0.028  $\pm$  0.014 & 0.040  $\pm$  0.032 & - & - & - - \\
176 & - & 17:45:40.3477 & -29:00:16.0054 & 12.73 & 9.49 & 106 $\times$ 79(117) & 0.074  $\pm$  0.013 & 0.241  $\pm$  0.056 & - & - & - - \\
177 & - & 17:45:41.0086 & -29:00:27.9045 & 12.73 & 22.91 & 348 $\times$ 212(25) & 0.059  $\pm$  0.009 & 1.130  $\pm$  0.189 & - & - & - - \\
178 & - & 17:45:40.3590 & -29:00:15.9615 & 12.82 & 13.24 & 88 $\times$ 0(46) & 0.048  $\pm$  0.014 & 0.084  $\pm$  0.036 & - & - & - - \\
179 & - & 17:45:40.8487 & -29:00:20.7756 & 12.89 & 15.01 & 172 $\times$ 111(138) & 0.064  $\pm$  0.013 & 0.375  $\pm$  0.089 & - & - & - - \\
180 & - & 17:45:41.0156 & -29:00:26.4324 & 12.92 & 15.38 & 115 $\times$ 55(25) & 0.052  $\pm$  0.014 & 0.138  $\pm$  0.048 & - & - & - - \\
181 & - & 17:45:39.0811 & -29:00:24.9056 & 12.95 & 9.40 & 240 $\times$ 0(33) & 0.132  $\pm$  0.013 & 0.493  $\pm$  0.062 & - & - & - - \\
182 & - & 17:45:40.9475 & -29:00:33.2946 & 13.02 & 13.35 & 296 $\times$ 189(10) & 0.105  $\pm$  0.011 & 1.531  $\pm$  0.167 & - & - & - - \\
183 & - & 17:45:40.8918 & -29:00:21.4103 & 13.03 & 13.63 & 184 $\times$ 140(97) & 0.068 $\pm$ 0.013 & 0.517 $\pm$ 0.112 & - & 0.50 $\pm$ 0.026 & - a \\
184 & - & 17:45:41.0282 & -29:00:26.9501 & 13.03 & 36.77 & 661 $\times$ 253(162) & 0.047  $\pm$  0.006 & 1.962  $\pm$  0.272 & - & - & - - \\
185 & IRS 5 & 17:45:40.6956 & -29:00:18.2766 & 13.05 & 0.84 & 81 $\times$ 59(171) & 0.834 $\pm$ 0.014 & 1.823 $\pm$ 0.041 & 39.93 $\pm$ 1.986 & 372.26 $\pm$ 19.543 & 4 - \\
186 & - & 17:45:39.6933 & -29:00:15.8245 & 13.05 & 18.29 & 138 $\times$ 60(147) & 0.047  $\pm$  0.013 & 0.152  $\pm$  0.056 & - & - & - - \\
187 & - & 17:45:41.0380 & -29:00:29.0877 & 13.15 & 31.11 & 100 $\times$ 18(30) & 0.023  $\pm$  0.014 & 0.044  $\pm$  0.037 & - & - & - - \\
188 & - & 17:45:41.0381 & -29:00:29.0851 & 13.16 & 32.15 & 109 $\times$ 23(31) & 0.023  $\pm$  0.014 & 0.048  $\pm$  0.040 & - & - & - - \\
189 & - & 17:45:40.4343 & -29:00:15.9412 & 13.19 & 20.95 & 178 $\times$ 97(36) & 0.047  $\pm$  0.013 & 0.258  $\pm$  0.084 & - & - & - - \\
190 & - & 17:45:40.8112 & -29:00:36.6198 & 13.26 & 4.02 & 212 $\times$ 145(11) & 0.314  $\pm$  0.013 & 2.682  $\pm$  0.124 & - & - & - - \\
191 & - & 17:45:39.6948 & -29:00:15.5601 & 13.30 & 13.57 & 126 $\times$ 88(133) & 0.058  $\pm$  0.013 & 0.225  $\pm$  0.064 & - & - & - - \\
192 & - & 17:45:40.4536 & -29:00:15.7828 & 13.44 & 9.66 & 153 $\times$ 123(131) & 0.094  $\pm$  0.013 & 0.543  $\pm$  0.088 & - & - & - - \\
193 & - & 17:45:40.4519 & -29:00:15.6462 & 13.56 & 17.36 & 243 $\times$ 182(168) & 0.075  $\pm$  0.012 & 0.883  $\pm$  0.152 & - & - & - - \\
194 & - & 17:45:39.7399 & -29:00:15.0726 & 13.57 & 12.17 & 266 $\times$ 46(66) & 0.067  $\pm$  0.013 & 0.418  $\pm$  0.094 & - & - & - - \\
195 & - & 17:45:39.1642 & -29:00:35.3859 & 13.60 & 16.27 & 520 $\times$ 337(84) & 0.057  $\pm$  0.006 & 2.500  $\pm$  0.279 & - & - & - - \\
196 & - & 17:45:41.0786 & -29:00:25.8958 & 13.82 & 6.16 & 149 $\times$ 35(174) & 0.162  $\pm$  0.014 & 0.399  $\pm$  0.045 & - & - & - - \\
197 & - & 17:45:41.0871 & -29:00:26.5116 & 13.85 & 11.44 & 177 $\times$ 54(7) & 0.098  $\pm$  0.013 & 0.336  $\pm$  0.058 & - & - & - - \\
198 & - & 17:45:41.0382 & -29:00:32.6689 & 13.90 & 14.17 & 129 $\times$ 60(32) & 0.058  $\pm$  0.013 & 0.180  $\pm$  0.053 & - & - & - - \\
199 & - & 17:45:38.9886 & -29:00:30.2981 & 13.95 & 23.68 & 336 $\times$ 71(159) & 0.077  $\pm$  0.013 & 0.567  $\pm$  0.109 & - & - & - - \\
200 & - & 17:45:41.0846 & -29:00:30.8325 & 14.00 & 10.54 & 166 $\times$ 23(34) & 0.089  $\pm$  0.014 & 0.263  $\pm$  0.051 & - & - & - - \\
201 & - & 17:45:40.4442 & -29:00:15.0899 & 14.03 & 16.66 & 211 $\times$ 80(149) & 0.069  $\pm$  0.013 & 0.375  $\pm$  0.084 & - & - & - - \\
202 & - & 17:45:40.4329 & -29:00:14.9786 & 14.08 & 19.20 & 387 $\times$ 100(115) & 0.053  $\pm$  0.012 & 0.652  $\pm$  0.155 & - & - & - - \\
203 & - & 17:45:39.0967 & -29:00:34.9493 & 14.14 & 21.34 & 254 $\times$ 87(161) & 0.067  $\pm$  0.013 & 0.443  $\pm$  0.099 & - & - & - - \\
204 & - & 17:45:41.1040 & -29:00:25.6805 & 14.18 & 4.71 & 166 $\times$ 41(28) & 0.209  $\pm$  0.013 & 0.664  $\pm$  0.054 & - & - & - - \\
205 & - & 17:45:41.1232 & -29:00:29.1250 & 14.27 & 23.23 & 115 $\times$ 70(175) & 0.036  $\pm$  0.014 & 0.106  $\pm$  0.052 & - & - & - - \\
206 & - & 17:45:41.1232 & -29:00:29.1241 & 14.27 & 23.63 & 119 $\times$ 78(173) & 0.036  $\pm$  0.013 & 0.117  $\pm$  0.056 & - & - & - - \\
207 & n40 & 17:45:41.1055 & -29:00:30.8395 & 14.27 & 13.37 & 216 $\times$ 89(31) & 0.088 $\pm$ 0.013 & 0.526 $\pm$ 0.091 & 0.25 $\pm$ 0.013 & 0.24 $\pm$ 0.016 & - - \\
208 & - & 17:45:40.0734 & -29:00:13.7723 & 14.30 & 9.08 & 203 $\times$ 51(128) & 0.098  $\pm$  0.013 & 0.453  $\pm$  0.073 & - & - & - - \\
209 & n41 & 17:45:41.0769 & -29:00:23.4960 & 14.37 & 5.83 & 117 $\times$ 55(155) & 0.138 $\pm$ 0.014 & 0.370 $\pm$ 0.048 & 0.27 $\pm$ 0.014 & 0.33 $\pm$ 0.018 & - - \\
210 & - & 17:45:39.0759 & -29:00:35.0124 & 14.41 & 31.37 & 91 $\times$ 24(156) & 0.023  $\pm$  0.014 & 0.040  $\pm$  0.035 & - & - & - - \\
211 & - & 17:45:41.1302 & -29:00:25.1992 & 14.61 & 25.13 & 113 $\times$ 28(20) & 0.032  $\pm$  0.014 & 0.065  $\pm$  0.039 & - & - & - - \\
212 & - & 17:45:41.1337 & -29:00:25.4370 & 14.61 & 12.35 & 90 $\times$ 43(50) & 0.052  $\pm$  0.014 & 0.117  $\pm$  0.042 & - & - & - - \\
213 & - & 17:45:41.0843 & -29:00:33.1266 & 14.62 & 12.68 & 271 $\times$ 112(165) & 0.121  $\pm$  0.013 & 1.022  $\pm$  0.122 & - & - & - - \\
214 & - & 17:45:41.0653 & -29:00:33.8645 & 14.67 & 4.29 & 192 $\times$ 70(15) & 0.271  $\pm$  0.013 & 1.190  $\pm$  0.070 & - & - & - - \\
215 & - & 17:45:41.1259 & -29:00:24.6722 & 14.67 & 6.59 & 116 $\times$ 33(168) & 0.128  $\pm$  0.014 & 0.264  $\pm$  0.040 & - & - & - - \\
216 & - & 17:45:41.1535 & -29:00:26.5113 & 14.71 & 10.50 & 211 $\times$ 63(168) & 0.120  $\pm$  0.013 & 0.533  $\pm$  0.071 & - & - & - - \\
217 & - & 17:45:41.1492 & -29:00:30.1266 & 14.72 & 24.79 & 97 $\times$ 36(161) & 0.030  $\pm$  0.014 & 0.059  $\pm$  0.038 & - & - & - - \\
218 & - & 17:45:41.1023 & -29:00:23.2924 & 14.75 & 43.55 & 344 $\times$ 63(13) & 0.044  $\pm$  0.013 & 0.305  $\pm$  0.103 & - & - & - - \\
219 & - & 17:45:41.1655 & -29:00:30.1966 & 14.94 & 31.38 & 87 $\times$ 46(170) & 0.023  $\pm$  0.014 & 0.045  $\pm$  0.039 & - & - & - - \\
220 & - & 17:45:41.1724 & -29:00:26.5239 & 14.96 & 20.30 & 207 $\times$ 33(12) & 0.062  $\pm$  0.013 & 0.201  $\pm$  0.055 & - & - & - - \\
221 & - & 17:45:41.1554 & -29:00:24.9905 & 14.97 & 9.91 & 194 $\times$ 33(22) & 0.115  $\pm$  0.013 & 0.377  $\pm$  0.056 & - & - & - - \\
222 & - & 17:45:41.1653 & -29:00:25.1825 & 15.06 & 12.59 & 198 $\times$ 87(9) & 0.095  $\pm$  0.013 & 0.500  $\pm$  0.081 & - & - & - - \\
223 & - & 17:45:40.0401 & -29:00:12.9664 & 15.10 & 12.21 & 226 $\times$ 112(78) & 0.067  $\pm$  0.013 & 0.534  $\pm$  0.116 & - & - & - - \\
224 & - & 17:45:39.1890 & -29:00:38.2955 & 15.12 & 20.37 & 141 $\times$ 94(125) & 0.039  $\pm$  0.013 & 0.177  $\pm$  0.072 & - & - & - - \\
225 & - & 17:45:41.1790 & -29:00:25.6167 & 15.16 & 25.20 & 267 $\times$ 42(130) & 0.043  $\pm$  0.013 & 0.242  $\pm$  0.086 & - & - & - - \\
226 & n42 & 17:45:41.0923 & -29:00:34.3266 & 15.18 & 5.38 & 123 $\times$ 50(88) & 0.108 $\pm$ 0.013 & 0.351 $\pm$ 0.056 & 0.02 $\pm$ 0.004 & 0.12 $\pm$ 0.012 & - - \\
227 & - & 17:45:41.1862 & -29:00:30.4889 & 15.25 & 20.83 & 177 $\times$ 125(14) & 0.052  $\pm$  0.013 & 0.335  $\pm$  0.097 & - & - & - - \\
228 & - & 17:45:41.1769 & -29:00:31.3529 & 15.29 & 4.59 & 99 $\times$ 53(10) & 0.167  $\pm$  0.014 & 0.383  $\pm$  0.043 & - & - & - - \\
229 & - & 17:45:39.8997 & -29:00:12.7857 & 15.39 & 15.36 & 181 $\times$ 163(133) & 0.070  $\pm$  0.013 & 0.584  $\pm$  0.121 & - & - & - - \\
230 & - & 17:45:41.2218 & -29:00:26.2774 & 15.63 & 24.91 & 217 $\times$ 82(6) & 0.052  $\pm$  0.013 & 0.284  $\pm$  0.083 & - & - & - - \\
231 & - & 17:45:40.1043 & -29:00:12.3756 & 15.72 & 9.92 & 67 $\times$ 0(125) & 0.055  $\pm$  0.014 & 0.065  $\pm$  0.028 & - & - & - - \\
232 & - & 17:45:41.1192 & -29:00:34.8645 & 15.72 & 13.62 & 483 $\times$ 168(10) & 0.136  $\pm$  0.009 & 2.829  $\pm$  0.196 & - & - & - - \\
233 & - & 17:45:39.6412 & -29:00:13.2314 & 15.73 & 5.57 & 263 $\times$ 82(98) & 0.125  $\pm$  0.013 & 0.973  $\pm$  0.114 & - & - & - - \\
234 & - & 17:45:41.0937 & -29:00:20.5394 & 15.76 & 33.52 & 181 $\times$ 85(179) & 0.034  $\pm$  0.013 & 0.160  $\pm$  0.075 & - & - & - - \\
235 & n43 & 17:45:40.4774 & -29:00:13.3632 & 15.79 & 24.70 & 196 $\times$ 122(50) & 0.040 $\pm$ 0.013 & 0.288 $\pm$ 0.106 & 23.80 $\pm$ 1.205 & 43.22 $\pm$ 2.269 & - - \\
236 & - & 17:45:40.4777 & -29:00:13.3583 & 15.80 & 24.97 & 262 $\times$ 136(66) & 0.040  $\pm$  0.013 & 0.418  $\pm$  0.144 & - & - & - - \\
237 & - & 17:45:41.0718 & -29:00:36.1862 & 15.80 & 12.91 & 457 $\times$ 347(150) & 0.095  $\pm$  0.007 & 3.756  $\pm$  0.265 & - & - & - - \\
238 & - & 17:45:41.1233 & -29:00:35.0698 & 15.86 & 6.49 & 172 $\times$ 0(27) & 0.158  $\pm$  0.014 & 0.404  $\pm$  0.046 & - & - & - - \\
239 & - & 17:45:41.2339 & -29:00:30.8441 & 15.93 & 11.33 & 242 $\times$ 77(45) & 0.098  $\pm$  0.013 & 0.629  $\pm$  0.096 & - & - & - - \\
240 & - & 17:45:41.2613 & -29:00:27.4785 & 16.05 & 16.12 & 111 $\times$ 53(170) & 0.051  $\pm$  0.014 & 0.124  $\pm$  0.045 & - & - & - - \\
241 & - & 17:45:41.2584 & -29:00:25.6813 & 16.18 & 13.86 & 215 $\times$ 64(158) & 0.089  $\pm$  0.013 & 0.413  $\pm$  0.073 & - & - & - - \\
242 & - & 17:45:41.2625 & -29:00:26.1201 & 16.18 & 14.51 & 111 $\times$ 44(141) & 0.050  $\pm$  0.014 & 0.125  $\pm$  0.045 & - & - & - - \\
243 & - & 17:45:41.2612 & -29:00:25.4845 & 16.25 & 5.51 & 88 $\times$ 60(153) & 0.128  $\pm$  0.014 & 0.303  $\pm$  0.044 & - & - & - - \\
244 & - & 17:45:41.2571 & -29:00:25.0159 & 16.28 & 8.86 & 208 $\times$ 0(147) & 0.126  $\pm$  0.014 & 0.361  $\pm$  0.050 & - & - & - - \\
245 & - & 17:45:41.2769 & -29:00:26.0669 & 16.37 & 15.63 & 93 $\times$ 0(143) & 0.043  $\pm$  0.014 & 0.067  $\pm$  0.033 & - & - & - - \\
246 & - & 17:45:38.8232 & -29:00:23.9115 & 16.47 & 17.78 & 157 $\times$ 91(43) & 0.050  $\pm$  0.013 & 0.237  $\pm$  0.075 & - & - & - - \\
247 & - & 17:45:41.2790 & -29:00:25.5622 & 16.47 & 11.69 & 158 $\times$ 37(135) & 0.070  $\pm$  0.013 & 0.228  $\pm$  0.055 & - & - & - - \\
248 & - & 17:45:41.2756 & -29:00:25.2239 & 16.48 & 17.39 & 250 $\times$ 39(4) & 0.086  $\pm$  0.013 & 0.338  $\pm$  0.064 & - & - & - - \\
249 & - & 17:45:41.2935 & -29:00:25.2157 & 16.71 & 28.60 & 288 $\times$ 46(134) & 0.043  $\pm$  0.013 & 0.261  $\pm$  0.091 & - & - & - - \\
250 & - & 17:45:41.2080 & -29:00:34.6999 & 16.72 & 12.44 & 150 $\times$ 24(134) & 0.064  $\pm$  0.014 & 0.185  $\pm$  0.051 & - & - & - - \\
251 & - & 17:45:41.3025 & -29:00:25.5407 & 16.78 & 18.19 & 196 $\times$ 66(142) & 0.057  $\pm$  0.013 & 0.265  $\pm$  0.074 & - & - & - - \\
252 & - & 17:45:39.0071 & -29:00:18.0328 & 16.84 & 27.97 & 340 $\times$ 122(11) & 0.064  $\pm$  0.012 & 0.720  $\pm$  0.149 & - & - & - - \\
253 & - & 17:45:38.8514 & -29:00:34.6366 & 16.90 & 32.01 & 266 $\times$ 26(151) & 0.044  $\pm$  0.013 & 0.187  $\pm$  0.068 & - & - & - - \\
254 & - & 17:45:40.7495 & -29:00:13.9038 & 16.96 & 10.84 & 383 $\times$ 258(118) & 0.097  $\pm$  0.008 & 2.463  $\pm$  0.215 & - & - & - - \\
255 & - & 17:45:41.2210 & -29:00:34.9159 & 16.96 & 7.13 & 55 $\times$ 33(11) & 0.087  $\pm$  0.014 & 0.127  $\pm$  0.032 & - & - & - - \\
256 & - & 17:45:41.3178 & -29:00:25.2774 & 17.02 & 30.85 & 261 $\times$ 67(174) & 0.050  $\pm$  0.013 & 0.270  $\pm$  0.084 & - & - & - - \\
257 & - & 17:45:39.7111 & -29:00:11.4971 & 17.12 & 15.44 & 159 $\times$ 70(74) & 0.043  $\pm$  0.013 & 0.196  $\pm$  0.072 & - & - & - - \\
258 & - & 17:45:41.3240 & -29:00:24.9570 & 17.15 & 6.88 & 24 $\times$ 0(3) & 0.064  $\pm$  0.015 & 0.058  $\pm$  0.024 & - & - & - - \\
259 & - & 17:45:41.2972 & -29:00:32.7643 & 17.17 & 11.04 & 156 $\times$ 93(33) & 0.085  $\pm$  0.013 & 0.400  $\pm$  0.074 & - & - & - - \\
260 & - & 17:45:41.3409 & -29:00:26.1492 & 17.20 & 22.75 & 152 $\times$ 0(152) & 0.041  $\pm$  0.014 & 0.082  $\pm$  0.039 & - & - & - - \\
261 & n44 & 17:45:39.7365 & -29:00:11.3046 & 17.23 & 7.82 & 324 $\times$ 118(97) & 0.099 $\pm$ 0.012 & 1.160 $\pm$ 0.152 & 0.16 $\pm$ 0.011 & 0.27 $\pm$ 0.018 & - - \\
262 & - & 17:45:39.3426 & -29:00:42.7085 & 17.25 & 7.62 & 217 $\times$ 24(115) & 0.095 $\pm$ 0.013 & 0.437  $\pm$  0.073 & 0.05 $\pm$ 0.003 & - & - b \\
263 & - & 17:45:41.3483 & -29:00:25.5177 & 17.37 & 16.52 & 164 $\times$ 43(3) & 0.065  $\pm$  0.014 & 0.186  $\pm$  0.050 & - & - & - - \\
264 & - & 17:45:39.3112 & -29:00:42.6938 & 17.46 & 11.26 & 403 $\times$ 375(167) & 0.104  $\pm$  0.007 & 3.916  $\pm$  0.259 & - & - & - - \\
265 & - & 17:45:41.3648 & -29:00:26.3373 & 17.49 & 28.58 & 250 $\times$ 0(142) & 0.043  $\pm$  0.014 & 0.117  $\pm$  0.049 & - & - & - - \\
266 & - & 17:45:38.8057 & -29:00:21.1645 & 17.58 & 14.10 & 118 $\times$ 82(50) & 0.053  $\pm$  0.013 & 0.191  $\pm$  0.060 & - & - & - - \\
267 & - & 17:45:41.3607 & -29:00:24.6881 & 17.67 & 7.90 & - & 0.055  $\pm$  0.015 & 0.037  $\pm$  0.019 & - & - & - - \\
268 & - & 17:45:41.3704 & -29:00:25.2665 & 17.70 & 11.58 & 61 $\times$ 23(2) & 0.056  $\pm$  0.014 & 0.076  $\pm$  0.030 & - & - & - - \\
269 & - & 17:45:38.8066 & -29:00:20.6206 & 17.79 & 15.78 & 168 $\times$ 107(35) & 0.062  $\pm$  0.013 & 0.344  $\pm$  0.085 & - & - & - - \\
270 & - & 17:45:39.2095 & -29:00:13.9728 & 17.80 & 18.67 & 133 $\times$ 91(12) & 0.048  $\pm$  0.013 & 0.190  $\pm$  0.065 & - & - & - - \\
271 & - & 17:45:38.9316 & -29:00:38.8503 & 18.08 & 11.97 & 170 $\times$ 121(8) & 0.089 $\pm$ 0.013 & 0.538  $\pm$  0.091 & 0.05 $\pm$ 0.003 & - & - b \\
272 & - & 17:45:39.0639 & -29:00:40.9216 & 18.13 & 14.91 & 237 $\times$ 64(168) & 0.093  $\pm$  0.013 & 0.458  $\pm$  0.077 & - & - & - - \\
273 & - & 17:45:41.3908 & -29:00:24.1526 & 18.17 & 18.63 & 184 $\times$ 87(15) & 0.060  $\pm$  0.013 & 0.299  $\pm$  0.077 & - & - & - - \\
274 & - & 17:45:38.7984 & -29:00:36.1891 & 18.18 & 22.13 & 210 $\times$ 102(4) & 0.057  $\pm$  0.013 & 0.358  $\pm$  0.094 & - & - & - - \\
275 & - & 17:45:41.4132 & -29:00:24.9639 & 18.30 & 27.23 & 227 $\times$ 119(154) & 0.047  $\pm$  0.013 & 0.359  $\pm$  0.113 & - & - & - - \\
276 & - & 17:45:38.7295 & -29:00:21.2475 & 18.47 & 17.47 & 122 $\times$ 65(37) & 0.045  $\pm$  0.013 & 0.142  $\pm$  0.054 & - & - & - - \\
277 & - & 17:45:41.4179 & -29:00:23.9070 & 18.57 & 42.25 & 185 $\times$ 92(173) & 0.027  $\pm$  0.013 & 0.138  $\pm$  0.080 & - & - & - - \\
278 & - & 17:45:38.6543 & -29:00:23.4128 & 18.74 & 19.64 & 302 $\times$ 73(7) & 0.088  $\pm$  0.013 & 0.588  $\pm$  0.099 & - & - & - - \\
279 & - & 17:45:38.7615 & -29:00:19.4502 & 18.84 & 15.84 & 61 $\times$ 41(93) & 0.036  $\pm$  0.014 & 0.066  $\pm$  0.037 & - & - & - - \\
280 & - & 17:45:41.4322 & -29:00:23.3200 & 18.89 & 41.46 & 272 $\times$ 116(134) & 0.030  $\pm$  0.013 & 0.277  $\pm$  0.133 & - & - & - - \\
281 & - & 17:45:41.4879 & -29:00:27.0501 & 19.04 & 28.21 & 525 $\times$ 140(169) & 0.074  $\pm$  0.009 & 1.418  $\pm$  0.189 & - & - & - - \\
282 & - & 17:45:41.3694 & -29:00:20.3179 & 19.10 & 17.07 & 169 $\times$ 29(119) & 0.041  $\pm$  0.013 & 0.150  $\pm$  0.061 & - & - & - - \\
283 & - & 17:45:41.3320 & -29:00:19.1094 & 19.19 & 17.87 & 120 $\times$ 0(137) & 0.041  $\pm$  0.014 & 0.089  $\pm$  0.042 & - & - & - - \\
284 & - & 17:45:41.4605 & -29:00:22.7906 & 19.39 & 26.27 & 337 $\times$ 155(155) & 0.058  $\pm$  0.011 & 0.804  $\pm$  0.164 & - & - & - - \\
285 & - & 17:45:38.6289 & -29:00:34.3247 & 19.52 & 12.00 & 193 $\times$ 75(129) & 0.075  $\pm$  0.013 & 0.388  $\pm$  0.080 & - & - & - - \\
286 & - & 17:45:41.4983 & -29:00:23.4224 & 19.71 & 33.76 & 311 $\times$ 114(163) & 0.051  $\pm$  0.013 & 0.500  $\pm$  0.140 & - & - & - - \\
287 & n45 & 17:45:41.5337 & -29:00:26.0544 & 19.72 & 26.84 & 181 $\times$ 103(54) & 0.034 $\pm$ 0.013 & 0.201 $\pm$ 0.091 & 0.14 $\pm$ 0.008 & 0.12 $\pm$ 0.008 & - - \\
288 & - & 17:45:38.6881 & -29:00:19.1668 & 19.82 & 23.82 & 244 $\times$ 169(57) & 0.046  $\pm$  0.012 & 0.529  $\pm$  0.150 & - & - & - - \\
289 & - & 17:45:41.5089 & -29:00:23.0871 & 19.92 & 18.00 & 143 $\times$ 45(14) & 0.053  $\pm$  0.014 & 0.144  $\pm$  0.049 & - & - & - - \\
290 & - & 17:45:41.4152 & -29:00:19.4268 & 20.02 & 34.72 & 209 $\times$ 74(50) & 0.028  $\pm$  0.013 & 0.153  $\pm$  0.085 & - & - & - - \\
291 & - & 17:45:41.4263 & -29:00:19.6998 & 20.04 & 12.41 & 124 $\times$ 0(60) & 0.042  $\pm$  0.015 & 0.000  $\pm$  0.002 & - & - & - - \\
292 & - & 17:45:38.5017 & -29:00:27.0224 & 20.19 & 19.61 & 68 $\times$ 0(169) & 0.034  $\pm$  0.014 & 0.042  $\pm$  0.029 & - & - & - - \\
293 & - & 17:45:38.8490 & -29:00:15.2240 & 20.21 & 9.25 & 116 $\times$ 95(78) & 0.080  $\pm$  0.013 & 0.313  $\pm$  0.064 & - & - & - - \\
294 & - & 17:45:41.5584 & -29:00:31.6049 & 20.25 & 9.42 & 232 $\times$ 65(57) & 0.097  $\pm$  0.013 & 0.574  $\pm$  0.090 & - & - & - - \\
295 & - & 17:45:41.4205 & -29:00:19.0474 & 20.25 & 31.72 & 175 $\times$ 33(145) & 0.031  $\pm$  0.013 & 0.098  $\pm$  0.055 & - & - & - - \\
296 & - & 17:45:41.5332 & -29:00:22.5257 & 20.38 & 13.58 & 142 $\times$ 73(178) & 0.070  $\pm$  0.013 & 0.246  $\pm$  0.059 & - & - & - - \\
297 & - & 17:45:41.2010 & -29:00:14.3333 & 20.53 & 10.14 & 126 $\times$ 45(133) & 0.072  $\pm$  0.014 & 0.206  $\pm$  0.050 & - & - & - - \\
298 & - & 17:45:41.4398 & -29:00:18.3367 & 20.80 & 68.39 & 293 $\times$ 99(177) & 0.025  $\pm$  0.013 & 0.202  $\pm$  0.119 & - & - & - - \\
299 & - & 17:45:41.5803 & -29:00:22.3161 & 21.03 & 15.00 & 181 $\times$ 74(110) & 0.048  $\pm$  0.013 & 0.245  $\pm$  0.080 & - & - & - - \\
300 & - & 17:45:41.4498 & -29:00:18.0416 & 21.06 & 17.46 & 196 $\times$ 105(115) & 0.049  $\pm$  0.013 & 0.324  $\pm$  0.099 & - & - & - - \\
301 & - & 17:45:38.6411 & -29:00:17.5982 & 21.11 & 17.81 & 395 $\times$ 338(48) & 0.063  $\pm$  0.007 & 2.132  $\pm$  0.245 & - & - & - - \\
302 & - & 17:45:39.9124 & -29:00:49.3133 & 21.31 & 9.83 & 149 $\times$ 127(126) & 0.092  $\pm$  0.013 & 0.533  $\pm$  0.089 & - & - & - - \\
303 & - & 17:45:38.5850 & -29:00:37.7890 & 21.40 & 31.64 & 370 $\times$ 74(136) & 0.050  $\pm$  0.013 & 0.459  $\pm$  0.132 & - & - & - - \\
304 & - & 17:45:41.6042 & -29:00:21.8618 & 21.46 & 12.82 & 102 $\times$ 53(12) & 0.060  $\pm$  0.014 & 0.141  $\pm$  0.043 & - & - & - - \\
305 & - & 17:45:38.8451 & -29:00:13.3551 & 21.48 & 19.04 & 228 $\times$ 76(13) & 0.070  $\pm$  0.013 & 0.379  $\pm$  0.083 & - & - & - - \\
306 & - & 17:45:41.6864 & -29:00:19.8473 & 23.13 & 16.92 & 187 $\times$ 57(32) & 0.061 $\pm$ 0.013 & 0.251 $\pm$ 0.066 & - & 0.33 $\pm$ 0.017 & - a \\
307 & - & 17:45:41.7140 & -29:00:19.2837 & 23.67 & 9.00 & 166 $\times$ 125(137) & 0.107  $\pm$  0.013 & 0.666  $\pm$  0.094 & - & - & - - \\
308 & - & 17:45:38.2814 & -29:00:21.9344 & 23.85 & 20.57 & 198 $\times$ 56(19) & 0.057  $\pm$  0.013 & 0.229  $\pm$  0.066 & - & - & - - \\
309 & - & 17:45:41.8513 & -29:00:30.2155 & 23.88 & 19.05 & 487 $\times$ 371(94) & 0.052  $\pm$  0.006 & 2.357  $\pm$  0.283 & - & - & - - \\
310 & - & 17:45:38.2666 & -29:00:21.6713 & 24.11 & 26.77 & 158 $\times$ 56(169) & 0.038  $\pm$  0.013 & 0.123  $\pm$  0.055 & - & - & - - \\
311 & - & 17:45:41.8561 & -29:00:23.6745 & 24.25 & 25.51 & 296 $\times$ 106(25) & 0.062  $\pm$  0.013 & 0.559  $\pm$  0.129 & - & - & - - \\
312 & - & 17:45:41.7261 & -29:00:17.8341 & 24.39 & 28.63 & 208 $\times$ 70(170) & 0.044  $\pm$  0.013 & 0.204  $\pm$  0.074 & - & - & - - \\
313 & - & 17:45:41.4663 & -29:00:11.6292 & 24.92 & 8.68 & 4 $\times$ 0(9) & 0.056  $\pm$  0.015 & 0.049  $\pm$  0.023 & - & - & - - \\
314 & - & 17:45:41.8096 & -29:00:17.9834 & 25.33 & 24.39 & 306 $\times$ 109(128) & 0.051  $\pm$  0.013 & 0.510  $\pm$  0.142 & - & - & - - \\
315 & - & 17:45:37.8554 & -29:00:26.1322 & 28.70 & 21.10 & 225 $\times$ 47(9) & 0.064  $\pm$  0.013 & 0.254  $\pm$  0.065 & - & - & - - \\
316 & - & 17:45:40.1361 & -28:59:58.8907 & 29.21 & 9.45 & - & 0.043  $\pm$  0.015 & 0.005  $\pm$  0.007 & - & - & - - \\
317 & - & 17:45:40.1236 & -28:59:58.1925 & 29.90 & 9.61 & 119 $\times$ 88(70) & 0.075  $\pm$  0.013 & 0.290  $\pm$  0.063 & - & - & - - \\
318 & - & 17:45:38.8944 & -29:00:57.8753 & 33.37 & 11.38 & 246 $\times$ 88(146) & 0.112  $\pm$  0.013 & 0.749  $\pm$  0.100 & - & - & - - \\-
\enddata
\tablecomments{\textbf{References}: 
1 -  Viehmann et al. (2005); 2 - Eckart et al. (2004); 3 - Paumard et al. (2006) and 4 - Moultaka et al. (2009). 
\textbf{Comments}: a  - Detected only in $L'$; b - Detected only in $K_s$ and faint source.}
\end{deluxetable}
\end{document}